\documentclass[13pt]{article}
\usepackage{epsf}
\usepackage{color}
\usepackage{framed}
\usepackage{amsmath,amsfonts,amsbsy,amssymb}
\numberwithin{equation}{section}
\usepackage{indentfirst}
\usepackage{mathtools}
\usepackage{arydshln} 
\usepackage{framed}

\newcommand{\fsl}[1]{\ensuremath{\mathrlap{\not{\phantom{#1}}}#1}}

\newcommand{\nn}{\nonumber}

\def\be{\begin{equation}}
\def\ee{\end{equation}}
\def\bse{\begin{subequations}}
\def\ese{\end{subequations}}
\def\bal{\begin{align}}
\def\ealn{\end{align}}
\def\tr{\text{tr}}
\def\bs{\boldsymbol}

\begin{document}

\begin{titlepage}

\def\slash#1{{\rlap{$#1$} \thinspace/}}

\begin{flushright} 

\end{flushright} 

\vspace{0.1cm}

\begin{Large}
\begin{center}

{\bf   Deformation of  Matrix Geometry  via  
  Landau Level Evolution }
\end{center}
\end{Large}

\vspace{1.0cm}

\begin{center}
{\bf Kazuki Hasebe}   \\ 

\vspace{0.5cm} 
\it{
National Institute of Technology, Sendai College,  
Ayashi, Sendai, 989-3128, Japan} \\
\vspace{0.2cm} 
\it{Department of Physics, University of Hong Kong,  Pokfulam Road, Hong 
Kong, China 
} \\

\vspace{0.6cm} 

{\sf
khasebe@sendai-nct.ac.jp} 

{\sf
hasebe@hku.hk} 

\vspace{0.6cm} 

{\today} 

\end{center}

\vspace{1.0cm}

\begin{abstract}
\noindent

\baselineskip=18pt

We propose a scheme for the construction of deformed matrix geometries using Landau models. The Landau models are practically useful tools to extract  matrix geometries.  The level projection method however cannot be applied straightforwardly  to the Landau models on  deformed manifolds, as  
they  do not generally exhibit degenerate energy levels. We overcome this problem by exploiting the idea of spectral flow. Taking a symmetric matrix geometry as a reference point of the spectral flow, we evolve the matrix geometry by deforming the Landau model. In this process, unitarity is automatically preserved. 
 The explicit matrix realization of the coordinates is derived mechanically  even for a non-perturbative deformation.  We clarify  basic properties of the deformed matrix geometries through a concrete analysis of the non-relativistic and relativistic Landau models on expanding two-sphere and elongating ellipsoid.  The obtained ellipsoidal matrix geometries show  behaviors quantitatively different in each Landau level, but  qualitatively similar to their classical counterpart. We also numerically investigate the differences between the ellipsoidal matrix geometry and the fuzzy ellipsoid.  

\end{abstract}

\end{titlepage}

\newpage 

\tableofcontents

\newpage 

\section{Introduction}

Non-commutative geometry is a mathematical framework for describing quantum space-time and is  a framework in progress, where many interesting ideas and  methods are being developed and applied. The idea of non-commutative geometry is important in many branches of physics, not only in particle physics \cite{Douglas-Nekrasov-2001,hep-th/9801182,Karabali-Nair-Daemi-2004, Hasebe-2010, Ho-Matsuo-2016} but also in condensed matter physics \cite{Girvin-Jach-1984,  Ezawa-Tsitsishvili-Hasebe-2003, Hasebe-2004, Neupertetal2012, Esienneetal2012, Shiozaki-Fujimoto-2013,  Hasebe-2014-1, Hasebe-2014-2, Hasebe-2017, Langmann-Ryu-Shiozaki-2024}, quantum information \cite{Hasebe-2024-1, Torma-2023}, 
and conformal field theory \cite{Zhu-Han-Huffman-Hofmann-He-2022, Cuomo-Komargodski-Mezei-RavivMoshe-2022, Cuomo-Fuente-Monin-Pirskhalava-Rattazzi-2018, Cuomo-Delacretaz-Mehta-2021}.  
One of the best known examples is the fuzzy sphere \cite{Hoppe1982, madore1992}, which realizes a classical solution of  Matrix models and the  geometry of the  Landau models.  Since the late 1990s,  the research on the  construction of fuzzy manifolds  has progressed significantly. In this development,  fuzzy manifolds corresponding to symmetric manifolds are established,  such as  fuzzy $\mathbb{C}P^n$ \cite{Alexanian-Balachandran-Immirzi-Ydri-2002, Balachandran-Dolan-Lee-Martin-Oconner-2002}, higher dimensional fuzzy spheres \cite{hep-th/9602115, Castelino-Lee-Taylor-1997, Ho-Ramgoolam-2002, Ramgoolam2002}, fuzzy hyperboloids \cite{Ho-Li-2000, Hasebe-2012} and their supersymmetric versions \cite{hep-th/9507074, math-ph/9804013, Hasebe-2011}. It has also been shown that their geometries can be realized in the Landau models,  
 for   $\mathbb{C}P^n$ \cite{Karabali-Nair-2002, Karabali-Nair-2006}, $S^n$ \cite{Hasebe-2016, Nair-Daemi-2004, Hasebe-2018, Zhang-Hu-2001, Hasebe-Kimura-2003, Ishiki-Matsumoto-Muraki-2018,  Hasebe-2020, Hasebe-2021,  Hasebe-2023-1}, their supersymmetric versions \cite{Hasebe-Kimura-2005, arXiv:hep-th/0311159}, and $T^n$ \cite{Adachi-Ishiki-Matsumoto-Saito-2020, Adachi-Ishiki-Kanno-2022}. 

Having explored the fuzzification of such symmetric manifolds, a natural direction to proceed is to consider the fuzzification of   manifolds with less  or no symmetries. 
There have been several early works already in this direction. To the best of author's knowledge,  the fuzzification of the 2D Riemannian surfaces with multiple genus has been investigated in Refs.\cite{Arnlind-Bordemann-Hofer-Hoppe-Shimada-2009-1, Arnlind-Bordemann-Hofer-Hoppe-Shimada-2009-2} based on  deformed algebra. For mathematical aspects of the classical geometry and its quantization,  the readers may consult with Refs.\cite{Klimek-Lesniewski-1992-1, Klimek-Lesniewski-1992-2, Bordemann-Meinrenken-Schlichenmaier-1993, Ma-Marinescu-2008}. 
Another important direction will be the time evolution of  fuzzy spaces.  An expanding fuzzy sphere has been analyzed in Refs.\cite{Sasakura-2004-1, Sasakura-2004-2}, where  the matrix size of the fuzzy space changes under the evolution,   and this evolution is  non-unitary   
An  expanding universe was also studied recently in the context of the type IIB matrix model  \cite{Hirasawa-et-al-2024, Hatakerama-et-al-2020}. The semi-classical analysis of the evolving non-commutative spacetime has  been performed in 
\cite{Steinacker-2018-1, Steinacker-2018-2, Stern-Xu-2018} based on the non-compact fuzzy spaces \cite{Hasebe-2012}.      
While the deformed fuzzy geometry and the time evolution of the fuzzy geometry are different in a strict sense, they are closely related as the time evolution usually accompanies with  the deformation of a space. It is also reported that the change of the entanglement entropy of a fuzzy space is perturbatively evaluated by deforming the base-space and the background field  of  the Landau model \cite{Nair-2020-2, Nair-2022}.

 In this work, we propose a scheme to construct non-symmetric matrix geometries in the perspective of  the evolution of matrix geometries. 
Recently, the author proposed to generate matrix geometries using the Landau models \cite{Hasebe-2023-1, Hasebe-2016}.\footnote{The Boerezin-Toeplitz quantization, which has been successfully applied to the matrix geometries \cite{Ishiki-Matsumoto-Muraki-2018, Nair-2020, Adachi-Ishiki-Matsumoto-Saito-2020, Matsuura-Tsuchiya-2020}, corresponds to the zeroth Landau level projection. In this sense, the present  method includes  the Berezin-Toeplitz quantization as a special case.  } Interestingly, it was shown that the   higher Landau levels exhibit  unique  matrix geometries distinct  from that of the lowest Landau level \cite{Hasebe-2023-1}. We combine this new scheme with the idea of spectral flow for deforming the matrix geometries. 
We adopt a symmetric matrix geometry as a reference point from which the spectral flow starts  and trace the evolution of the matrix geometry from it to a non-symmetric manifold of interest along the spectral flow.    
This method naturally realizes the deformation of non-commutative space associated with the time evolution. While this scheme is very simple and intuitive,  it has several unique advantages.  First,  the derivation of the matrix coordinates, which  are very important in practical applications, is straightforward.  
 In the former approaches based on deformed algebra, the concrete matrix realizations to satisfy the deformed algebra are not feasible to find in general.   
Second, since the spectral flow is a concept irrelevant to perturbation, we can discuss non-perturbative deformations of the matrix geometries.   
Lastly,  the unitarity of the non-commutative space is automatically guaranteed  (without resorting to  scenarios such as multi-universe \cite{Sasakura-2004-1}), as we can trace the evolution of the states by the spectral flow.       
To demonstrate  this scheme, we consider two concrete evolutions of the fuzzy two-sphere. The first is the expanding fuzzy sphere, which keeps the $SU(2)$ symmetry during the evolution. The other is the deformation to  ellipsoidal matrix geometry, where the $SU(2)$ symmetry is no longer guaranteed and the Landau level degeneracy is lifted.  In this process, the spectral flow is crucial for tracing the evolution of the Landau level eigenstates.   

The paper is organized as follows. In Sec.\ref{defmat-spectralf}, we present the idea of the deformation of matrix geometries using the spectral flow. In Sec.\ref{subsec:matfuzzsphe}, we review the fuzzy sphere obtained from the Landau level projection. The matrix geometry of the expanding sphere is discussed. Section \ref{sec:ellipmagf} presents the classical geometry and  the algebra on the ellipsoid. We introduce a magnetic field to be compatible with the geometry of the ellipsoid.   In Sec.\ref{sec:non-rellan}, we numerically solve the non-relativistic Landau problem on the ellipsoid and derive the spectral flow with respect to the deformation from sphere to ellipsoid.  The corresponding matrix geometries are derived.     
Their relativistic versions are investigated in Sec.\ref{sec:rella}.  
In Sec.\ref{sec:diffmatelfuz}, we numerically analyze basic features of the ellipsoidal matrix geometries in comparison with the commutative ellipsoids  and the fuzzy ellipsoids.
Section \ref{sec:summary} is devoted to summary and discussion.

\section{Prescription for the deformation of  matrix geometry }\label{defmat-spectralf}

In this section, we propose a scheme for the deformation of  matrix geometries from the perspective of the spectral flow.  

\subsection{Matrix geometry for $G/H$}\label{subsec:mategh}

To generate the matrix geometry of $\mathcal{M}\simeq G/H$, we use the Landau model with gauge symmetry $H$ on 
$\mathcal{M}$ \cite{Hasebe-2023-1, Hasebe-2016}.   We  apply the 
 level projection to a Landau level and extract the matrix geometry:
\be
(X_i)_{mn}=\langle \psi_m|x_i|\psi_n\rangle, \label{formulacoma}
\ee
where $|\psi_m\rangle$ denote the degenerate eigenstates in the Landau level. The size of the matrix coordinates $X_i$ is equal to the degeneracy of the Landau level. This scheme is practically useful for  deriving the explicit form of the matrix coordinates. However, when we consider to apply this procedure to  non-symmetric manifolds, 
     we  immediately encounter a difficulty.  
 Since  the energy levels of the Landau model on a non-symmetric manifold do $\it{not}$  accommodate  degeneracies in  general,   we simply do not know how to identify the energy levels to which the level projection is applied.

\subsection{Matrix geometry deformation via the spectral flow}\label{subsec:matgesp}

\begin{figure}[tbph]
\center
\includegraphics*[width=110mm]{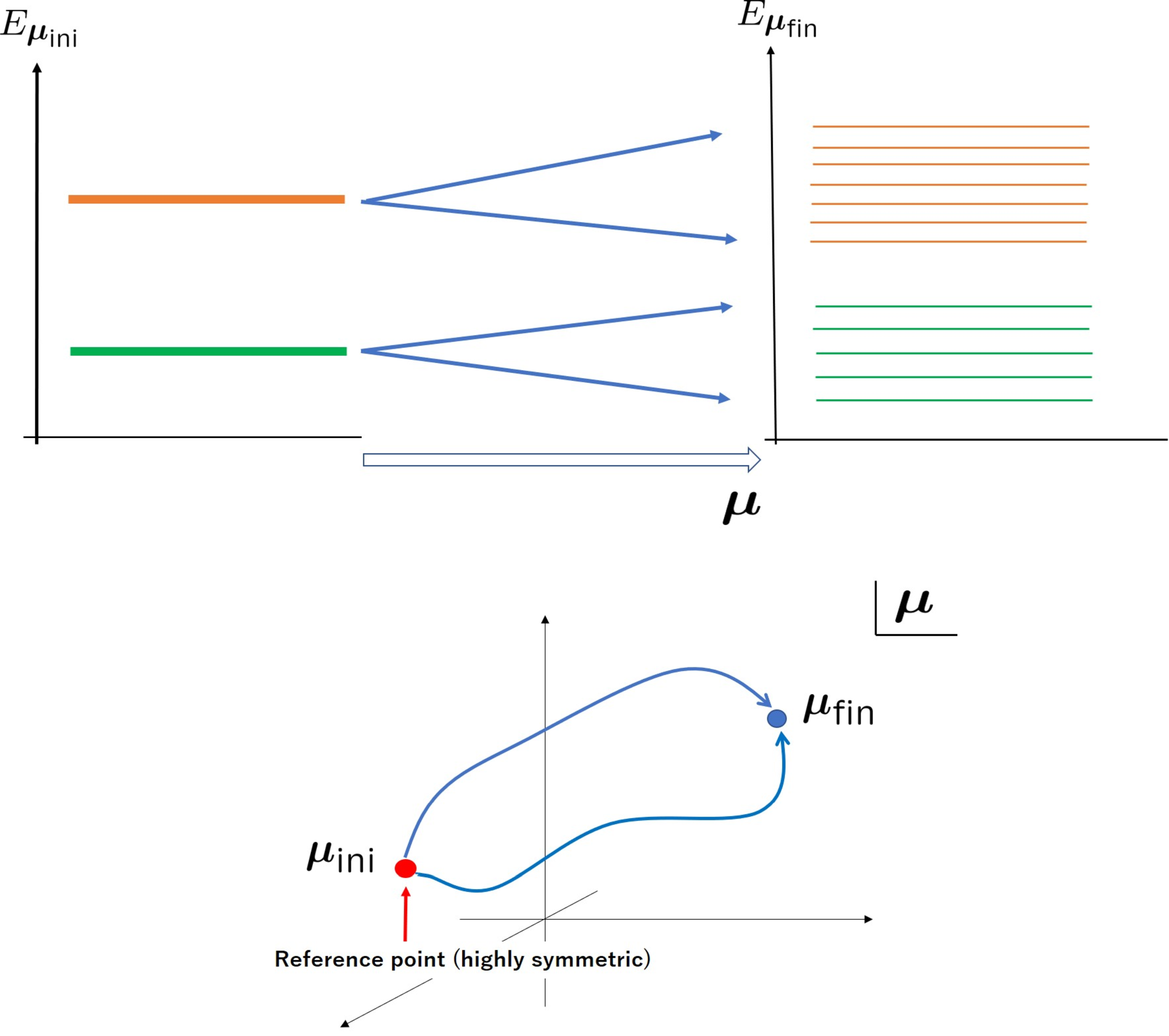}
\caption{Evolution of the Landau levels. Upper: the degeneracy of the Landau level on a symmetric manifold  (left) is lifted by a deformation (right).  Lower: there are an infinite number of paths in the parameter space  that connect the initial value of the deformation parameter and the final value. (In the  figure, only two paths are depicted.)}
\label{def.fig}
\end{figure}

Any non-symmetric manifold can be obtained from a continuous deformation of a symmetric manifold (of the same topology). We extend this basic picture to derive  non-symmetric  matrix geometry. To translate this classical picture in the language of quantum mechanics, we consider two Landau models on a symmetric manifold and  non-symmetric manifold of interest. 
We  introduce a deformation parameter $\bs{\mu}$ to connect these two manifolds.\footnote{$\bs{\mu}$ can be a set of parameters.}  
 We construct a parameter-dependent Hamiltonian $H^{(\bs{\mu})}$ that interpolates the two Landau Hamiltonians on the symmetric manifold and the non-symmetric manifold. 
  Take  $\bs{\mu}=\bs{\mu}_{\text{ini}}$ for the initial symmetric manifold and another value $\bs{\mu}_{\text{fin}}$ for the final non-symmetric manifold. 
The  Landau Hamiltonian on the symmetric manifold is denoted by $H^{(\bs{\mu}_{\text{ini}})}$, and its eigenvalues  $E^{(\bs{\mu}_{\text{ini}})}$ signify the Landau level with degeneracy. Similarly, we denote the  Landau Hamiltonian on the non-symmetric manifold as $H^{(\bs{\mu}_{\text{fin}})}$ and its eigenvalues as  $E^{(\bs{\mu}_{\text{fin}})}$, which  are no longer degenerate in general. 
We then solve the eigenvalue problem of the parameter dependent Hamiltonian, 
\be
H^{(\bs{\mu})}\psi^{(\bs{\mu})} =E^{(\bs{\mu})}\psi^{(\bs{\mu})}. 
\ee
Here, $E^{(\bs{\mu})}$ signifies the spectral flow of the eigen-energies from $E^{(\bs{\mu}_{\text{ini}})}$ to $E^{(\bs{\mu}_{\text{fin}})}$ (see the upper part of Fig.\ref{def.fig}). 
 Since the energy eigenstates are associated with the energy eigenvalues, 
 it is also possible to follow the evolution of the eigenstates from $\psi^{(\bs{\mu}_{\text{ini}})}$ to $\psi^{(\bs{\mu}_{\text{fin}})}$  by tracing the spectral flow. 
In this way, we can identify the eigenstates on the non-symmetric  manifold needed for the level projection (\ref{formulacoma}).  
Since the  degrees of freedom do not increase or decrease in the process of spectral flow,  the unitarity is guaranteed in this deformation process. Also note  that  the spectral flow itself is a concept irrelevant to perturbation and  we can access a  non-perturbative deformation of the matrix geometry.

As there are an infinite number of paths that connect $\bs{\mu}_{\text{ini}}$ and $\bs{\mu}_{\text{fin}}$ (the lower part of Fig.\ref{def.fig}),  
careful readers may wonder whether the evolved eigenstates will depend on the choice of the path and  the matrix coordinates also vary according to the chosen path.   
Obviously, the  energy spectrum of the deformed manifold does not depend on the path, however, the phases of the eigenstates do depend  on the path which they undergo. 
Consequently, the explicit form of the matrix coordinates  obtained  by the formula (\ref{formulacoma}) generally depends on the path chosen, but the difference of the eigenstates are only their phases\footnote{If the energy levels are degenerate, the (geometric) phase is not simple $U(1)$, but becomes a  Wilczek-Zee non-Abelian phase. Then, the difference of the eigenstates is accounted for by a unitary transformation.}, and the two matrix coordinates corresponding to  different paths  are related by a unitary transformation.  In other words,  the obtained matrix coordinates are unitarily equivalent,   and so the algebraic structure of the matrix geometry is independent of the choice of path.

\subsection{Matrix geometry evolution and unitarity}

We parametrize the path $\bs{\mu}$  as 
\be
\bs{\mu}=\bs{\mu}(t) ~~~~~~~(\bs{\mu}_{\text{ini}}:=\bs{\mu}(t=0)). 
\ee
The matrix coordinates are given by 
\be
X_i(t) =P_N(t)x_i P_N(t), \label{projecxa}
\ee
where $P_N(t)$ denotes the projection operator onto the subspace of the eigenstates evolving from the $N$th Landau level with degeneracy $d_N$:   
\be
P_N(t):= \sum_{n=1}^{d_N}|\psi_n(t)\rangle \langle \psi_n(t)|.
\ee
Assuming that $t$ represents the time parameter, $|\psi_n(t)\rangle$ satisfies  the time-dependent Schr\"odinger equation: 
\be
i\frac{d}{d t}|\psi_{n}(t)\rangle =H(t)|\psi_n(t)\rangle. \label{schropsi}
\ee
At $t=0$, $|\psi_n(0)\rangle$ denote the degenerate Landau level eigenstates: 
\be
H(0)|\psi_n(0)\rangle =E_N |\psi_n(0)\rangle~~~~~(n=1,2,\cdots, d_N).
\ee
For $t\neq 0$, the Landau level degeneracy is generally lifted, and   $|\psi_n(t)\rangle$ become the eigenstates with different energy levels.  
From (\ref{schropsi}), we can see that 
$P_N(t)$ satisfies 
\be
\frac{d}{dt} P_N(t) =i[P_N(t), H(t)]. 
\ee
A straightforward calculation tells\footnote{ 
Using the coordinates in the Heisenberg picture $h_i:= U(t)^{\dagger}x_i(t) U(t) ~~(U(t):=\text{T}e^{-i\int_0^t ds H(s)})$ and the  Heisenberg equations of motion 
\be
\frac{d}{dt}h_i(t) =U(t)^{\dagger}\biggl(\frac{dx_i(t)}{dt}-i[x_i(t), {H}(t)] \biggr)U(t), 
\ee
(\ref{timexa}) can be concisely represented as 
\be
\frac{dX_i(t)}{dt}-i[X_i(t), H(t)] = P_N(t) U(t)\frac{d{h_i}(t)}{dt} U(t)^{\dagger}P_N(t). \label{timexa2}
\ee
} 
\be
\frac{dX_i(t)}{dt}-i[X_i(t), H(t)]  =P_N(t)\biggl(\frac{dx_i(t)}{dt} -i [x_i(t), H(t)] \biggr)P_N(t) . \label{timexa}
\ee
Equation (\ref{timexa}) signifies the relation between the evolution of the matrix coordinates  and that of the classical coordinates. 
We can  derive the evolution of the matrix coordinates  by solving  the differential equation (\ref{timexa}) with given $H(t)$,  $P_N(t)$ and $x_i(t)$.  Note that   $\frac{d}{dt}X_i$  is not simply equal to the projection of $\frac{d}{dt}x_i$ ($\frac{dX_i(t)}{dt}\neq  P_N(t) \frac{dx_i(t)}{dt}P_N(t)$).   In the case $P_N=1$, $X_i$ is equal to $x_i$, and  Eq.(\ref{timexa}) becomes a trivial equation.  While Eq.(\ref{timexa}) determines the evolution of the matrix coordinates,  it is practical to use  the following formula to trace  the evolution of the matrix geometry: 
\be
(X_i(t))_{mn} =\langle \psi_m(t)|x_i(t)|\psi_n(t)\rangle .
\ee
Since we follow the unitary-evolution of the state-vector, the unitarity  of non-commutative space is $\it{necessarily}$ guaranteed  in the present scheme.

\section{Matrix geometry of  the fuzzy sphere }\label{subsec:matfuzzsphe}

In this section, we give a brief review of the fuzzy two-sphere that emerges in the $SO(3)$ Landau model. 
We also discuss the expanding fuzzy sphere to see how  the present scheme works. 

\subsection{Classical algebra of sphere-coordinates}

Let us start with the classical algebra on a two-sphere.  The two-sphere is homeomorphic to $\mathbb{C}P^1$,  which 
 is a K\"ahler manifold with  the K\"aher form 
\be
K =i\frac{2}{(1+|z|^2)^2} dz\wedge d\bar{z}. \label{kahlerforms2}
\ee
The complex coordinate $z$ is known as the inhomogeneous coordinate of $\mathbb{C}P^1$. 
We can parametrize $z$ as 
\be
z=\tan(\frac{\theta}{2})~e^{i\phi}~~~~~~(0\le \theta <\pi, ~~0\le \phi <2\pi). \label{stereoz}
\ee
With the coordinates on $S^2$, 
\be
x_1 =\sin\theta\cos\phi, ~~~~x_2 =\sin\theta\sin\phi, ~~~~x_3 =\cos\theta, \label{xscoopol}
\ee
we can find that the inhomogeneous coordinate $z$ (\ref{stereoz}) is equal to the stereographic coordinates on $\mathbb{C}\simeq \mathbb{R}^2$:  
\be
z=\frac{1}{1+x_3}(x_1+ix_2). 
\ee
The K\"ahler form (\ref{kahlerforms2})  is expressed as 
\be
K=\sin\theta~ d\theta \wedge d\phi, \label{kahang}
\ee
which is nothing but   the area element of $S^2$. 
Regarding the K\"ahler form as a symplectic form, we  consider the K\"aher manifold  as a symplectic manifold and  rewrite (\ref{kahang}) as the canonical form, $K=d\zeta\wedge d\phi$, where $\zeta$ and $\phi$ are the Darboux coordinates (canonical coordinates) given by 
\be
\zeta := \int d\theta ~\sin\theta =-\cos\theta. \label{zetaeq}
\ee
Using (\ref{kahang}),  we can represent the Poisson bracket on the $S^2$ as 
\be
\{f, g\} = \frac{\partial f}{\partial \zeta}\frac{\partial g}{\partial \phi}-\frac{\partial f}{\partial \phi}\frac{\partial g}{\partial \zeta}=
\frac{1}{\sin\theta} \biggl(\frac{\partial f}{\partial \theta}\frac{\partial g}{\partial \phi}-\frac{\partial f}{\partial \phi}\frac{\partial g}{\partial \theta}\biggr). \label{poissonbra}
\ee
Substituting (\ref{xscoopol}) to (\ref{poissonbra}), we can derive the classical algebra   of  the sphere-coordinates: 
\be
\{x_i, x_j\} =\sum_{k=1}^3 \epsilon_{ijk}x_k~~~~~~(\sum_{i=1}^3 x_ix_i=1). \label{s2poiss}
\ee

\subsection{$SO(3)$ Landau model and the fuzzy two-sphere}

\begin{figure}[tbph]
\center
\includegraphics*[width=140mm]{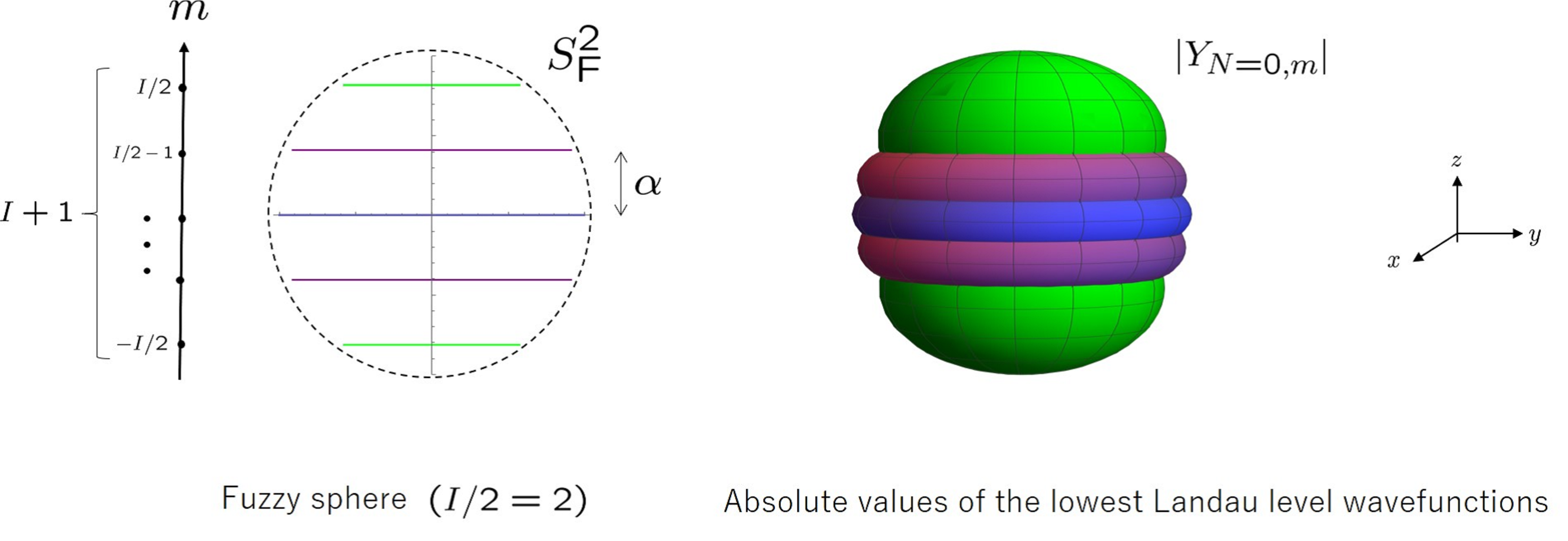}
\caption{Left: the matrix geometry of the fuzzy sphere (the dashed circle is added as a  guide for the eyes). The eigenvalues of ${X}_3/\alpha$  are plotted on the $m$-axis. The lengths of the latitudes correspond  the eigenvalues of $2\sqrt{{X}^2+{Y}^2}$.    Right: the distributions of the absolute values  of the monopole harmonics for $m=\pm 2$ (green), $\pm 1$ (purple), and  $0$ (blue). 
}
\label{fs2.fig}
\end{figure}

Next, we discuss a quantum mechanical counterpart of (\ref{s2poiss}) following \cite{Hasebe-2016}. 
On a sphere with unit radius, the $SO(3)$ Landau Hamiltonian  is given by \cite{Wu-Yang-1976, Haldane-1983}
\be
H=-\frac{1}{2M} \sum_{i=1}^3 (\partial_i+iA_i)^2|_{r=1}=\frac{1}{2M}\sum_{i=1}^3{\Lambda_{i}}^2=-\frac{1}{2M}\biggl(\partial_{\theta}^2 +\cot\theta \partial_{\theta} +\frac{1}{\sin^2\theta} (\partial_{\phi}+iA_{\phi})^2 \biggr) \label{landausph2}
\ee
where 
\be
\Lambda_i =-i\sum_{k=1}^3 \epsilon_{ijk}x_j (\partial_k +iA_k) .
\ee
In the following, we adopt the Schwinger gauge, \footnote{See Ref.\cite{Hasebe-2016} for instance.} where the gauge field  is represented as  
\be
A=-\frac{I}{2}\cos\theta d\phi,~~~~F=dA =\frac{I}{2}\sin\theta d\theta \wedge d\phi = \frac{1}{2}\epsilon_{ijk}B_k dx_i\wedge dx_j
\ee 
with 
\be
B_i=\frac{I}{2 r^3}x_i. 
\ee
Here $I$ denotes a quantized monopole change and is taken to be a positive integer henceforth. In the Schwinger gauge, the direction of the  gauge field is opposite in the north and south hemi-spheres,  $A_{\phi}(\theta) = -\frac{I}{2}\cos\theta =-A_{\phi}(\pi-\theta)$.  In other words, the gauge field has the mirror symmetry with respect to $z=0$. 
The magnetic field is perpendicular to the surface of the two-sphere and the configuration of the magnetic field  respects the  $SO(3)$ symmetry. The gauge field figuration  is thus compatible with the geometry of the base-manifold $S^2$.  
The Landau Hamiltonian (\ref{landausph2}) respects  the $SO(3)$ symmetry generated by the angular momentum operators 
\be
[H, L_i]=0, 
\ee
with  
\be
L_i:= \Lambda_i+r^2 B_i,
\ee
which satisfy the $su(2)$ algebra 
\be
[L_i, L_j] =i\sum_{k=1}^3 \epsilon_{ijk}L_k. 
\ee
Using a group theoretical method, it is not difficult to obtain the eigenvalues of $H$ (\ref{landausph2}) as 
\be
E_N =\frac{1}{2M} (N(N+1) +I(N+\frac{1}{2})), \label{spherelleiv}
\ee
where $N(=0,1,2,\cdots)$ denotes the Landau level index.  The degeneracy of the $N$th Landau level is given by 
 $2N+I+1$. The degenerate Landau level eigenstates are known as the monopole harmonics $Y_{N, m}(\theta, \phi)$ 
with $m$ being the quantum magnetic number: 
\be
 m=\overbrace{N+\frac{I}{2}, N+\frac{I}{2}-1, \cdots, -N-\frac{I}{2}}^{2N+I+1}. \label{rangem}
\ee
The monopole harmonics are normalized as  $\int_{S^2} d\theta d\phi \sin\theta ~Y_{N, m} (\theta, \phi)^*  Y_{N', m'}(\theta, \phi)=\delta_{NN'}\delta_{mm'}$. 
From the formula (\ref{formulacoma}), the matrix coordinates in the $N$th Landau level are derived as \cite{Hasebe-2016}
\be
 (X_i)_{mn}:=\int_{S^2} d\theta d\phi \sin\theta ~Y_{N, m} (\theta, \phi)^* x_i Y_{N, n}(\theta, \phi)=\alpha(I, N)(S_i^{(\frac{I}{2}+N)})_{mn}, \label{llspinmat} 
\ee
where 
\be
\alpha(I, N): = \frac{2I}{(I+2N)(I+2N+2)}. \label{ncscale}
\ee
The parameter  $\alpha(I, N)$ is referred to as the non-commutative scale that describes  the spacing between the adjacent states on the fuzzy sphere (see the left part of Fig.\ref{fs2.fig}). The  matrix size of $X_i$  is equal to the  Landau level degeneracy, $2N+I+1$. The coordinates are effectively represented by the $su(2)$ matrices with the spin index $J=\frac{I}{2}+N$ and   satisfy 
\be
\sum_{i=1}^3 X_iX_i =r(I,N)^2~\bs{1}_{I+2N+1}, ~~~~~~
[X_i, X_j] =i\alpha(I, N)~\sum_{k=1}^3\epsilon_{ijk}X_k,  \label{comxxx}
\ee
where 
\be
r(I, N):=\alpha(I, N)\cdot  \frac{\sqrt{(I+2N)(I+2N+2)}}{2} =\frac{I}{\sqrt{(I+2N)(I+2N+2)}}. 
\ee
Note that (\ref{comxxx}) realizes a natural quantum   counterpart of the classical relation (\ref{s2poiss}).  The geometry of the fuzzy sphere emerges in $\it{any}$ of the  Landau levels as in (\ref{comxxx}). The second relation of (\ref{comxxx}) implies that the matrix coordinates  act as the generators of the $SU(2)$ rotations for the fuzzy two-sphere. 
 The fuzzy sphere obviously has the $SU(2)$ symmetry 
\be
[\bs{X}^2, {X}_i] =0, 
\ee
which is a consequence of  the $SO(3)$ symmetry of the original Landau model.  
Since three ${X}_i$s are not commutative,  there is no concept of definite points for the fuzzy sphere due to the uncertainly relation of the coordinates.   Three matrix coordinates are not simultaneously diagonalized, but only one of ${X}_i$ is  diagonalized to represent the ``points'' on the fuzzy sphere. We take  this matrix coordinate as ${X}_3$, whose eigenvalues represent  the positions of the latitudes as shown in the left part of  Fig.\ref{fs2.fig}.  In other words, the latitudes, which are spaced by the non-commutative scale $\alpha$, denote the points constituting the fuzzy sphere.  
The right of Fig.\ref{fs2.fig} represents the absolute values of the  lowest Landau level eigenstates (the monopole harmonics of $N=0$). 
Obviously, there is one to one correspondence between the constituent ``points'' of the matrix geometry and the  Landau level eigenstates.  Then, the fuzzy sphere can be seen as a set  of the degenerate eigenstates of Landau level.

 The continuum limit corresponds to $I\rightarrow \infty$ in which the non-commutative scale (\ref{ncscale}) goes to zero $\alpha(I, N)\rightarrow 0$.   Indeed in this limit, the right-hand side of  the second equation of (\ref{comxxx}) vanishes, meaning that $X_i$ become  commutative coordinates. 
 Similarly, the first equation of (\ref{comxxx})   is reduced to the definition of a continuum sphere with unit radius ($r(I,N) ~\rightarrow~1$).

For the relativistic Landau model, the fuzzy spheres emerge again as the matrix geometry in each Landau level \cite{Hasebe-2016}. In particular, the matrix coordinates  in the zeroth Landau level are given by  
\be
X_i =\alpha(I-1)~S_i^{(\frac{I}{2}-\frac{1}{2})}~~~~\biggl(\alpha(I-1):= \alpha(I-1, N=0)=\frac{2}{I+1}\biggr), \label{zeromodexcoord}
\ee
which satisfy 
\be 
\sum_{i=1}^3 X_i X_i = \frac{I-1}{I+1}\bs{1}_I, ~~~~~~[X_i, X_j] =i\alpha(I-1)~\sum_{k=1}^3\epsilon_{ijk}X_k. \label{zeromoderad}
\ee 

\subsection{Expanding fuzzy sphere}\label{sec:evsphe}

To see how the present scheme  works, we consider an expanding process of the fuzzy two-sphere. The classical two-sphere with radius $\mu$ is given by 
\be
x^{(\mu)}x^{(\mu)}+y^{(\mu)}y^{(\mu)}+z^{(\mu)}z^{(\mu)}=\mu^2, \label{classicalexp}
\ee
or 
\be
x_i^{(\mu)}=\mu x_{i}. \label{classicalexp2}
\ee
Apparently, the $\mu$ plays the role of the deformation parameter of the expanding two-sphere.  Since the $SO(3)$ global  symmetry is preserved in the evolution process, the degeneracy of each Landau level is not lifted. We expect that the  matrix geometry  obtained will realize  an expanding fuzzy two-sphere. 
 Since the number of points of the fuzzy sphere is preserved  in the present scheme, the only possibility to realize the expanding fuzzy sphere is that the non-commutative length itself grows with the evolution.\footnote{The non-unitary evolution of the fuzzy space is discussed by Sasakura \cite{Sasakura-2004-1}. In the paper, the non-commutative scale is fixed and the dimensions of the irreducible representation change in the evolution process, and so  the evolution is non-unitary.  }

\begin{figure}[tbph]
\center
\includegraphics*[width=100mm]{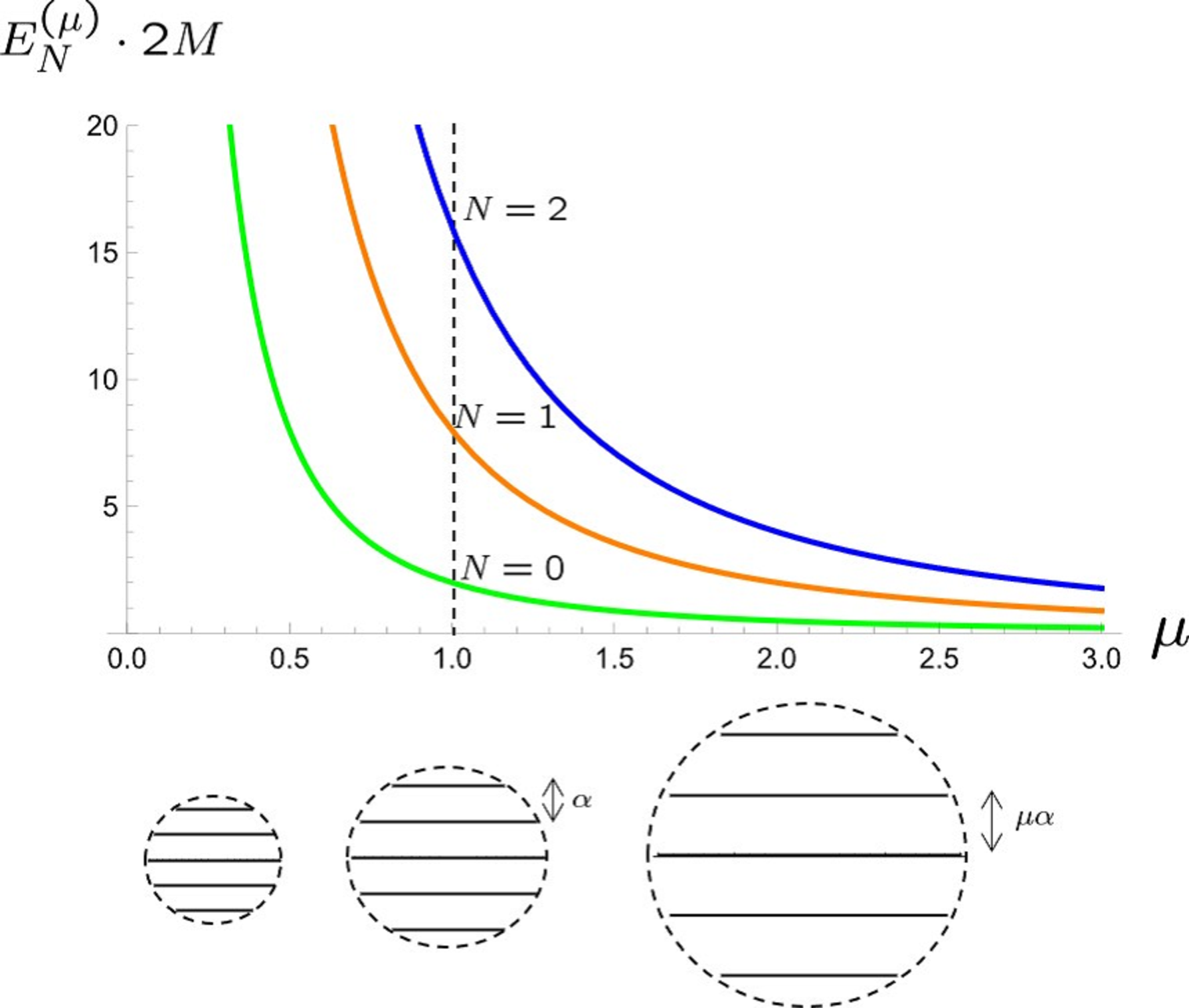}
\caption{The spectral flow of $E_{N}^{(\mu)}$ (\ref{muenesta}) for $I/2=2$ and $N=0,1,2$ (green, orange, blue).   The  degeneracy in each Landau level is not lifted.  }
\label{exsp.fig}
\end{figure}

We consider the Landau Hamiltonian on the sphere with radius $\mu$: 
\be
H^{(\mu)}= -\frac{1}{2M \mu^2}\biggl(\partial_{\theta}^2 +\cot\theta \partial_{\theta} +\frac{1}{\sin^2\theta} (\partial_{\phi}-iA_{\phi})^2 \biggr)=\frac{1}{\mu^2}H,
\ee
where $H$ is  (\ref{landausph2}). 
The energy eigenstates and eigenstates are readily obtained as  
\be
E_{N}^{(\mu)}=\frac{1}{\mu^2}E_N, ~~~~~Y_{N,n}^{(\mu)}=\frac{1}{\mu}Y_{N,m}(\theta, \phi), \label{muenesta}
\ee
where $E_N$ is given by (\ref{spherelleiv}) and $Y_{N,m}$ denote the monopole harmonics. Note that $Y_{N,n}^{(\mu)}$ are normalized on the surface of the two-sphere with radius $\mu$.  
See Fig.\ref{exsp.fig} for the spectral flow of $E^{(\mu)}$. The degeneracy in each Landau level is not lifted in the evolution process. 
We can easily derive the matrix coordinates  as 
\be
(X_i^{(\mu)})_{mn}=\int \mu^2 d\theta d\phi \sin\theta ~Y_{N,m}^{(\mu)}(\theta, \phi)^*~x^{(\mu)}_i~Y^{(\mu)}_{N,n}(\theta, \phi)=\mu (X_i)_{mn},  \label{mulxi}
\ee
which satisfy 
\be
\sum_{i=1}^3 X_i^{(\mu)}X_i^{(\mu)}=\mu^2 \sum_{i=1}^3 X_iX_i, ~~~~~[X_i^{(\mu)}, X_j^{(\mu)} ]=i\mu\alpha(I, N)\sum_{k=1}^3 \epsilon_{ijk}X_k^{(\mu)}. \label{musc2}
\ee
Here, $X_i$ are given by (\ref{llspinmat}). 
As anticipated, (\ref{mulxi}) shows that the expansion of the  fuzzy sphere geometry is in accordance with the behavior of the original   classical geometry (\ref{classicalexp2}).  The non-commutative parameter becomes $\mu\alpha(I,N)$ as shown in the second equation of (\ref{musc2}), and it grows as  the classical sphere expands. 
This simple demonstration confirms the plausibility of the present scheme.  

\section{Classical ellipsoidal geometry and magnetic field}\label{sec:ellipmagf}

\begin{figure}[tbph]
\center
\includegraphics*[width=180mm]{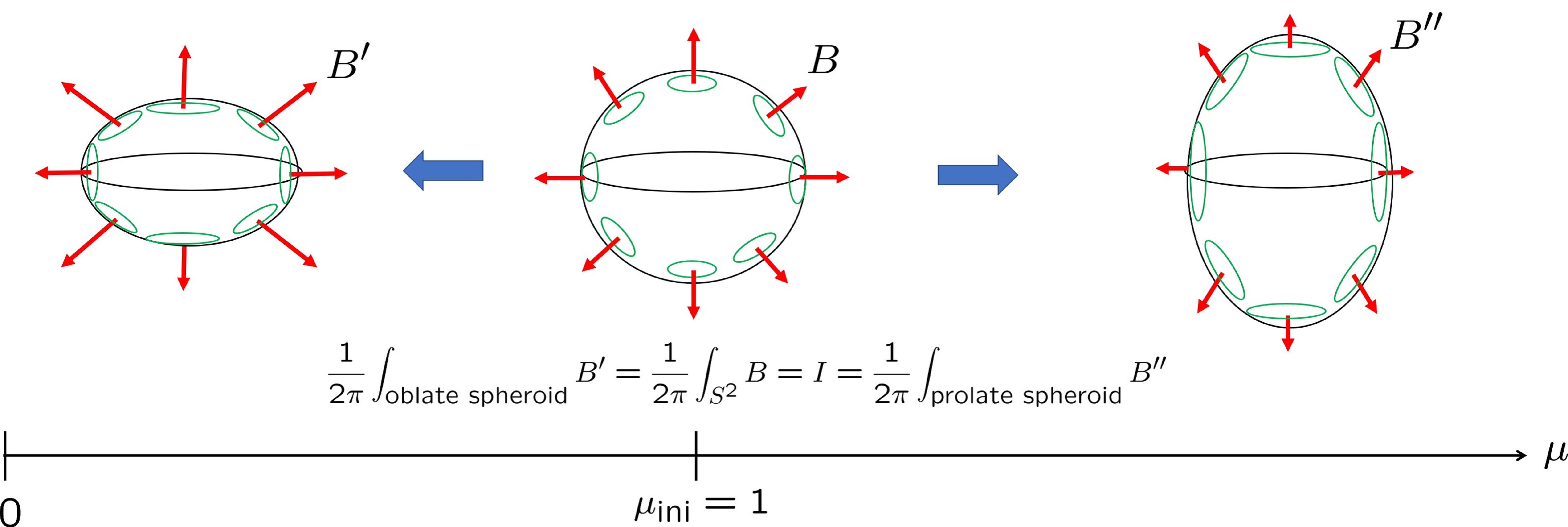}
\caption{Deformation from sphere to ellipsoids. The $\mu_{\text{ini}}=1$ is the reference point at which the two-sphere is realized.  The  $\mu<1$  represents an oblate spheroid (the limit $\mu \rightarrow 0$ corresponds to the  squashed sphere \cite{Andronache-Steinacker-2015}), 
while the $\mu>1$ denotes a prolate spheroid.  
 }
\label{deffromsphere.fig}
\end{figure}

Next,  we apply the present scheme for the derivation of  ellipsoidal matrix geometry. The classical evolution process is shown in Fig.\ref{deffromsphere.fig}. 
As the evolution process does not respect the $SU(2)$ symmetry, 
the degeneracy of each Landau level is lifted.  Tracing the spectral flow is crucial to identify the eigenstates.   
As a preparation,  this section introduces  basic geometries and algebraic structures of the classical ellipsoid.

\subsection{Geometric quantities }

We introduce the ellipsoid as 
\be
x^2+y^2+\biggl(\frac{z}{\mu}\biggr)^2=1, \label{defellipsoid}
\ee
where $\mu ~(>0)$ denotes the deformation parameter of the ellipsoid.  We call  the limit $\mu\rightarrow 1/0$  the sphere-limit/squashed-sphere-limit in this paper. 
We  parametrize  the ellipsoid in the ellipto-spherical coordinates \cite{Dassios-book-2012}: 
\be
x=\sin\theta\cos\phi, ~~~y=\sin\theta\sin\phi,~~z=\mu\cos\theta, 
\ee
where $\theta=[0, \pi]$ and $\phi=[0, 2\pi)$. 
The area element of the ellipsoid is given by (\ref{areaele})  
\be
d\Omega =\rho~\sin\theta d\theta\wedge d\phi=d\mathit{\Omega}_{\phi}(\theta)\wedge  d\phi, \label{areaelel}
\ee
where 
\be
\rho:=  \sqrt{\mu^2(x^2+y^2)+\frac{1}{\mu^2}z^2} =\sqrt{\cos^2\theta +\mu^2 \sin^2\theta}. 
\ee
As in the $S^2$ case, we will take the symplectic form as the area element  (\ref{areaelel}). 
Then, the  symplectic potential $\Omega$  is given by  
\be
\Omega =\biggl(\int \rho ~\sin\theta d\theta \biggr)\wedge d\phi . \label{sympcomp}
\ee
The integral is evaluated as 
\be
\int \rho~\sin\theta d\theta  =\mathit{\Omega}_{\phi}(\theta) +f(\phi) \label{sympcompint}
\ee
with $f(\phi)$ being an arbitrary function of $\phi$ and\footnote{
For $\mu<1$, we can represent  (\ref{firstw}) as   
\be
 \mathit{\Omega}_{\phi}(\theta) =-\frac{1}{2}( 1+\rho \cos\theta) +\frac{\mu^2}  {\sqrt{1-\mu^2}}\text{Arctanh} \biggl(  \frac{1-\rho}{\sqrt{1-\mu^2}}\frac{1}{1+\cos\theta} \biggr).
\ee
} 
\be
\mathit{\Omega}_{\phi}(\theta) =-\frac{1}{2}( 1+\rho \cos\theta) -\frac{\mu^2}  {\sqrt{\mu^2-1}}\text{Arctan} \biggl(  \frac{1-\rho}{\sqrt{\mu^2-1}}\frac{1}{1+\cos\theta} \biggr). \label{firstw}
\ee
In the limit $\mu~\rightarrow~1$, $\mathit{\Omega}_{\phi}(\theta)$ is reduced to $\zeta$ (\ref{zetaeq}): 
\be
\mathit{\Omega}_{\phi}(\theta) ~\rightarrow~\zeta(\theta)=-\cos\theta,   
\ee
which has the mirror symmetry with respect to $z=0$, $i.e.$, $\zeta(\pi-\theta)=-\zeta(\theta)$.
Meanwhile,  $\mathit{\Omega}_{\phi}(\theta)$ itself does not have the mirror symmetry: 
\be
\mathit{\Omega}_{\phi}(\theta) \neq -\mathit{\Omega}_{\phi}(\pi-\theta) ~~~~(\mu \neq 1).
\ee
Using the degrees of freedom $f(\phi)$,\footnote{ $\mathit{\Omega}_{\phi}(\theta)$ (\ref{firstw}) satisfies $\partial_{\theta}\mathit{\Omega}_{\phi}(\theta)=-\partial_{\theta}\mathit{\Omega}_{\phi}(\pi-\theta)$, and so  $\mathit{\Omega}_{\phi}(\theta)+\mathit{\Omega}_{\phi}(\pi-\theta)$ does not depend on $\theta$. Then, we  take $f(\phi)=-\frac{1}{2}(\mathit{\Omega}_{\phi}(\theta)+\mathit{\Omega}_{\phi}(\pi-\theta))$ in (\ref{sympcompint}) to derive (\ref{defnewsym}). } 
we introduce  the following symplectic potential:  
\be
\Omega_{\phi}(\theta) =\frac{1}{2}(\mathit{\Omega}_{\phi}(\theta)-\mathit{\Omega}_{\phi}(\pi-\theta)) \label{defnewsym}
\ee
with the mirror symmetry 
\be
\Omega_{\phi}(\pi-\theta)=-\Omega_{\phi}(\theta) . 
\ee
The explicit form of $\Omega_{\phi}(\theta)$ is given by\footnote{In the Cartesian coordinates, (\ref{schwga}) is given by 
\be
\Omega =\sum_{i=1}^3\Omega_i dx_i,
\ee
with 
\be
\Omega_i =\frac{1}{x^2+y^2}
\sum_{j=1}^3\epsilon_{ij3}x_j  \cdot\biggl( \frac{\rho}{2\mu} z +\frac{\mu^2}  {2\sqrt{\mu^2-1}}\biggl(\text{Arctan} \biggl( \frac{1-\rho}{\sqrt{\mu^2-1}}\frac{\mu}{\mu+z} \biggr)  - \text{Arctan} \biggl( \frac{1-\rho}{\sqrt{\mu^2-1}}\frac{\mu}{\mu-z} \biggr)   \biggr)\biggr). \label{omegaacar}
\ee
In the sphere-limit $\mu\rightarrow 1$, $\Omega_i ~\rightarrow~\frac{z}{\sqrt{x^2+y^2+z^2}~(x^2+y^2)}\sum_{j=1}^3 \epsilon_{ij3}x_j$, which is the Dirac monopole gauge field in the Schwinger gauge \cite{Hasebe-2016}. 
}
\be
\Omega_{\phi}(\theta) =-\frac{1}{2}\rho\cos\theta + \frac{\mu^2}  {2\sqrt{\mu^2-1}}\biggl(\text{Arctan} \biggl( \frac{1-\rho}{\sqrt{\mu^2-1}}\frac{1}{1-\cos\theta} \biggr)-  \text{Arctan} \biggl( \frac{1-\rho}{\sqrt{\mu^2-1}}\frac{1}{1+\cos\theta} \biggr)  \biggr) . \label{schwga}
\ee
It is straightforward to check  
\be
\partial_{\theta}\Omega_{\phi}(\theta) =\rho~\sin\theta \label{parome}
\ee
and 
\be
\Omega_{\phi}(\theta) ~\overset{\mu~\rightarrow ~1}\longrightarrow~-\cos\theta. \label{redomeg}
\ee
In the following, we adopt $\Omega_{\phi}(\theta)d\phi$ (\ref{schwga}) as our symplectic potential. For the sphere, the spin connection  coincides with the symplectic potential,$-\cos\theta d\phi$, while for the ellipsoid   they are different (compare (\ref{spinconel}) and (\ref{schwga})).

\subsection{Classical algebra of ellipsoid-coordinates}\label{subsec:classellip}

While the ellipsoid is obviously symmetric around the $SO(2)$ rotation around the $z$-axis, it also has a ``hidden'' $SO(3)$ symmetry:  When we rescale $z$ by $\mu z$, the ellipsoid is reduced a two-sphere to have the $SO(3)$ rotational symmetry. 
Indeed, the ellipsoid (\ref{defellipsoid}) is apparently symmetric with respect to the $SO(2)$ rotation around the $z$-axis generated by 
\be
L_z =-ix\partial_y +iy\partial_x=-i\partial_{\phi}, \label{mul3}
\ee
and also symmetric under the rotations generated by    
\be
L_x =-i\mu y\partial_{z} +i \frac{1}{\mu} z\partial_y=i\sin\phi\partial_{\theta}+i\cos\phi\cot\theta \partial_{\phi}, ~~L_y =-i\frac{1}{\mu} z\partial_{x} +i {\mu} x\partial_z=-i\cos\phi\partial_{\theta}+i\sin\phi\cot\theta \partial_{\phi}.
\label{mul12}
\ee
They obviously satisfy the $so(3)$ algebra,  
$[L_i, L_j]=i\sum_{k=1}^3 \epsilon_{ijk} L_k$, 
which implies that  the ellipsoid has the  ``$SO(3)$ symmetry''.  
Meanwhile, the classical algebra on  the ellipsoid only respects the $SO(2)$ symmetry  as we shall see below.    
With the symplectic form (\ref{sympcomp}),  
   $\Omega_{\phi}$ (\ref{schwga}) and $\phi$  signify the Darboux coordinates (canonical coordinates), and the Poisson bracket on the ellipsoid is represented as 
\be
\{f, g\}=\frac{\partial f}{\partial \Omega_{\phi}}\frac{\partial g}{\partial \phi}-\frac{\partial f}{\partial \phi}\frac{\partial g}{\partial \Omega_{\phi}} =\frac{1}{\sqrt{\cos^2\theta +\mu^2\sin^2\theta}~\sin\theta} \biggl(\frac{\partial f}{\partial \theta}\frac{\partial g}{\partial \phi}-\frac{\partial f}{\partial \phi}\frac{\partial g}{\partial \theta}\biggr). 
\ee
Then, we have 
\be
\{x, y\} =\frac{1}{\sqrt{ \mu^4 ({x}^2+{y}^2)  +{z}^2 }}z,  ~~~~\{y, z\} =\frac{\mu^2}{ \sqrt{ \mu^4 ({x}^2+{y}^2)  +{z}^2 }}x,  ~~~~\{z, x\} =\frac{\mu^2}{ \sqrt{ \mu^4 ({x}^2+{y}^2)  +{z}^2 }}y. \label{poisellip}
\ee
Equation (\ref{poisellip}) is expanded in terms of  $\delta \mu := \mu-1$ as 
\be
\{x, y\} \simeq z -\delta \mu \cdot z(2-{z}^2),  ~~~~\{y, z\} \simeq x+\delta \mu \cdot x {z}^2,   ~~~~\{z, x\} \simeq y+\delta \mu \cdot y {z}^2, 
\ee
where  ${x}^2+{y}^2=1-\frac{1}{\mu^2}{z}^2$ was used.  The algebra  is  highly nonlinear  unlike the case of  two-sphere $(\delta\mu=0)$. 
The Poisson bracket relations (\ref{poisellip}) are covariant under the $SO(2)$ rotation around the $z$-axis, but not under the  rotations generated by (\ref{mul12}).   
It is expected that the ellipsoidal matrix geometries will have the $U(1)$ symmetry rather than the $SU(2)$ symmetry, since they should realize  a quantum version of the classical algebra (\ref{poisellip}).

\subsection{Gauge field and magnetic field}

We introduce the gauge field to be proportional to the symplectic potential: 
\be
A =\frac{2\pi I}{S}\Omega_{\phi}(\theta) d\phi=\frac{2\pi I}{S}\sum_{i=1}^3 \Omega_i dx_i \label{gaugefiele}
\ee
where $S$ denotes the area of the ellipsoid (\ref{areaexelli}). 
The present gauge field has the mirror symmetry, and then we refer to the present gauge  as the Schwinger gauge.  In the sphere-limit, $A$ (\ref{gaugefiele}) is  reduced to the Dirac monopole gauge field,  $A~\rightarrow~-\frac{I}{2}\cos\theta ~d\phi$, while  in the squashed-sphere limit, $A~\rightarrow~-\frac{I}{2}\cos^2\theta~ d\phi$. 
The corresponding magnetic field is proportional to the symplectic form: 
\be
F=dA =\frac{2\pi I}{S}d\Omega=\frac{2\pi I}{S}\rho ~\sin\theta d\theta \wedge 
d\phi =F_{\theta\phi}d\theta\wedge d\phi, \label{ellpsfieldstrength}
\ee
where 
\be
F_{\theta\phi}=-F_{\phi\theta} =\partial_{\theta}A_{\phi}-\partial_{\phi}A_{\theta}=\frac{2\pi I}{S}\partial_{\theta}\Omega_{\phi}=  \frac{2\pi I}{S}\rho ~\sin\theta. \label{prefield}
\ee
Here, $I$ is  the first Chern number of the  monopole charge
\be
\frac{1}{2\pi}\int_{\text{ellipsoid}} F=I. \label{1stchern}
\ee
Notice that the field strength (\ref{prefield}) is invariant under the parallel transport on the ellipsoid,  
\be
\nabla_{\mu}F_{\nu\rho}=\partial_{\mu}F_{\nu\rho} -\Gamma_{~~\mu\nu}^{\lambda}F_{\lambda\rho} - \Gamma_{~~\mu\rho}^{\lambda}F_{\nu\lambda}=0. \label{nablaf0}
\ee
This means that the present magnetic field is uniform on the ellipsoid and compatible with the symmetry of the ellipsoid. It is important to take  the gauge field configuration  to be compatible with the geometry of the manifold, otherwise the  matrix geometry obtained will not have the same symmetry as the original classical manifold. 
Since the gauge field owes its origin to the geometry of the manifold (and indeed the gauge field is essentially the symplectic potential), the gauge field is not a fixed configuration, but it evolves along with the deformation of the ellipsoid.\footnote{In Ref.\cite{Matsuura-Tsuchiya-2020}, Matsuura and Tsuchiya analyzed the Berezin-Toeplitz quantization of the ellipsoid.    Their gauge field configuration is a $\it{fixed}$ Dirac monopole unlike the present  gauge field  (\ref{gaugefiele}). Then, their model is different from the present model.  }  

Let us regard this system as a 2D system embedded in $\mathbb{R}^3$. 
In the whole $\mathbb{R}^3$ space, there are an infinite number of  magnetic fields that  
yield (\ref{ellpsfieldstrength}) on the ellipsoid. 
Typical examples of such 3D magnetic fields may be given by        
\be
B_i:= \frac{2\pi I}{Sr^4} \sqrt{ 
 x^2+y^2+\frac{1}{\mu^4}z^2} ~x_i, \label{magcar}
\ee
and 
\be
B'_i := \frac{2\pi I}{S r^2}n_i, \label{orthomag}
\ee
where 
\be
r:= \sqrt{x^2+y^2+\frac{1}{\mu^2}z^2}, 
\ee
and $n_i$ denote the normal vector to the surface of the ellipsoid\footnote{The normal vector (\ref{normalvecell}) can be derived by the formula,   $\bs{n} =\frac{1}{|\frac{\partial\bs{x}}{\partial\theta} \times \frac{\partial\bs{x}}{\partial\phi}|}\frac{\partial\bs{x}}{\partial\theta} \times \frac{\partial\bs{x}}{\partial\phi}$. Also from Eq.(\ref{defellipsoid}),  we have  
\be
xdx +ydy +\frac{1}{\mu^2}z dz =0,  
\ee
and then  $(n_x, n_y, n_z) \propto (x,y, \frac{1}{\mu^2}z)$. }  
\be
(n_x, n_y, n_z) =\frac{1}{\rho}(
\mu x , ~
\mu y , ~
\frac{1}{\mu} z ). 
\label{normalvecell}
\ee
The $\bs{B}'$ (\ref{orthomag}) is obviously orthogonal to the surface of the ellipsoid.  
In both cases, they have the same component perpendicular to the ellipsoid: 
\be
B_{\perp}:=  \bs{B} \cdot \bs{n}=\bs{B}'\cdot  \bs{n} =\frac{2\pi I}{S}. \label{bana}
\ee
This is consistent with the uniformity of the magnetic field (\ref{nablaf0}).  
With (\ref{bana}),  we can simply reproduce (\ref{1stchern}) as 
\be
\frac{1}{2\pi}\int_{\text{ellipsoid}} B_{\perp}  d\Omega =I.
\ee
As we consider an electron confined on the surface of the ellipsoid, the electron  only feels the Lorentz force induced by the perpendicular component of the magnetic field. In other words,   these two magnetic fields affect the electron  in the exact same way.  Note, however, from the perspective of the existence of a vector potential in the 3D space,  $\bs{B}$   is more plausible  than  $\bs{B}'$.  This is because  the divergence of $\bs{B}'$ does not vanish ($\text{div} \bs{B}'\neq 0$), which implies that there is no gauge potential $\bs{A}'$ to satisfy    $\bs{B}'=\text{rot} \bs{A}'$.  
 Meanwhile,   the   $\bs{B}$ is divergence free (except for the origin), $\text{div} \bs{B}= 0$ ($\bs{x}\neq 0$), and a gauge potential $\bs{A}$ exists for the $\bs{B}$.    Indeed, $\bs{A}=\frac{2\pi I}{S}\bs{\Omega}$ (\ref{omegaacar}) gives  $\bs{B}$   by  $\text{rot} \bs{A}=\bs{B}$.    

\begin{figure}[tbph]
\center
\includegraphics*[width=140mm]{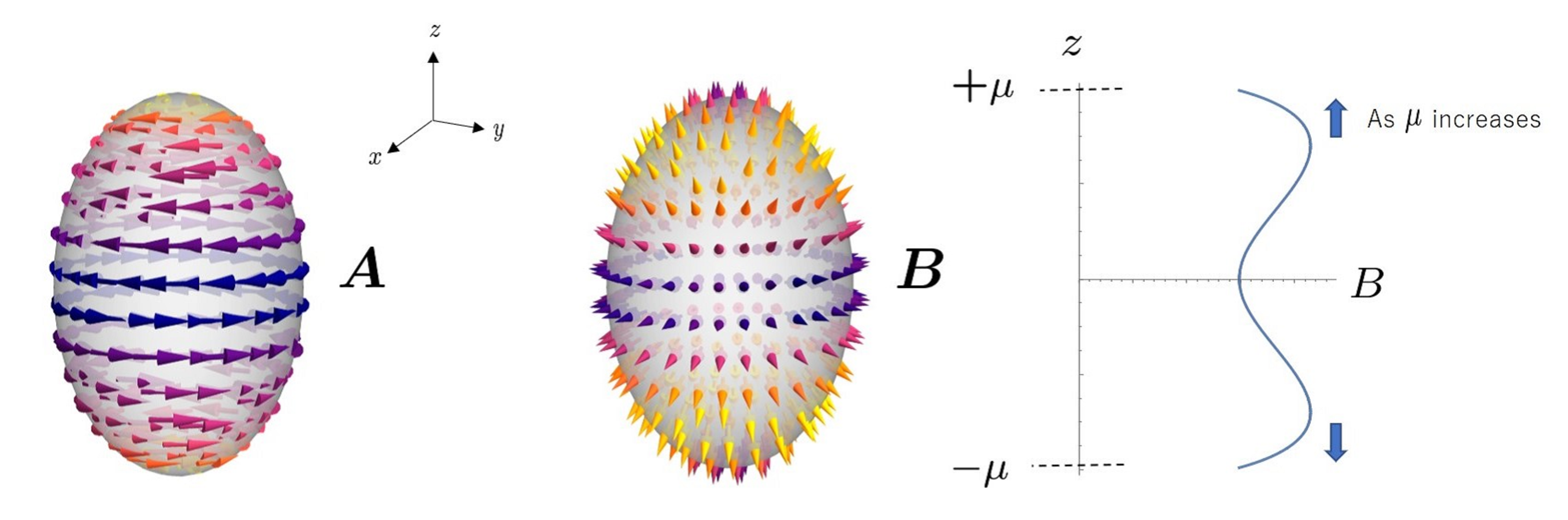}
\caption{Left: the distribution of the gauge field $\bs{A}$ on the ellipsoid (\ref{gaugefiele}). Middle: the distribution of the magnetic field $\bs{B}$  (\ref{magcar}). Right: the distribution of the magnitude $B$ (\ref{magmags}) with non-trivial dependence on $z$. The two peaks move to the directions of the blue arrows as $\mu$ increases.      }
\label{EntirePicture.fig}
\end{figure}

The magnetic field $\bs{B}$ (\ref{magcar}) has the singularity at the origin $x=y=z=0$, and the field  configuration is radial but  not spherically symmetric.
The distribution of the magnetic field (\ref{magcar}) is axially symmetric around the $z$-axis.   While the ellipsoid is transformed to a sphere by the scale transformation 
$z\rightarrow \mu z$ and has the ``$SO(3)$ symmetry'' (see Sec.\ref{subsec:classellip}),  $\bs{B}$ (\ref{magcar}) is not transformed to the magnetic field  of the Dirac monopole by this transformation. The existence of the magnetic field reduces the symmetry of the system from ``$SO(3)$'' 
to $SO(2)$. 
The magnitude of $\bs{B}$ is given by  
\be
B=\frac{2\pi I}{S r^4} \sqrt{\biggl(x^2+y^2+\frac{1}{\mu^4} z^2\biggr) \biggl(x^2+y^2+z^2\biggr)}.
\label{magmag}
\ee
On the surface of the ellipsoid $(r=1)$,  (\ref{magmag}) is reduced to 
\be
B|_{r=1}=\frac{2\pi I}{S}\sqrt{\biggl(1-\frac{1}{\mu^4} (\mu^2-1) z^2\biggr)\biggl(1+\frac{1}{\mu^2} (\mu^2-1) z^2\biggr) } . \label{magmags}
\ee
 In the sphere-limit $(\mu\rightarrow 1)$, the magnetic field becomes constant $B=\frac{I}{2\pi}$. 
 See Fig.\ref{EntirePicture.fig} for the configurations of $\bs{A}$ and $\bs{B}$.

\section{Non-relativistic Landau model and  matrix geometries}\label{sec:non-rellan}

In this section, we numerically solve the eigenvalue problem of the Landau model on the ellipsoid and  derive ellipsoidal  matrix geometry.

\subsection{Symmetry and eigenvalue problem on an ellipsoid}

From the general formula of the Laplace-Beltrami operator on the Riemannian manifold 
\be
\Delta=\frac{1}{\sqrt{g}} (\partial_{\mu}+iA_{\mu}) (\sqrt{g} g^{\mu\nu}(\partial_{\nu}+iA_{\nu})), 
\ee
we can easily construct the Landau Hamiltonian on the ellipsoid as 
\be
H^{(\mu)}=-\frac{1}{2M} \biggl(\frac{1}{\rho^2}\partial_{\theta}^2 +\cot\theta \frac{1}{\rho^4}\partial_{\theta} +\frac{1}{ \sin^2\theta} (\partial_{\phi}+iA_{\phi}(\theta))^2\biggr). \label{landauham} 
\ee
At $\mu=1$, (\ref{landauham}) is reduced to the Landau Hamiltonian on a two-sphere (\ref{landausph2}), which  respects the $SU(2)$ rotational symmetry. 
Recall that $\rho$ and $A_{\phi}$ depend on the deformation parameter $\mu$. 
The Landau Hamiltonian (\ref{landauham}) is commutative with 
\be
L_z =-ix\frac{\partial}{\partial y} +iy\frac{\partial}{\partial x}=-i\frac{\partial}{\partial \phi}, 
\ee
which means that this Hamiltonian has the $U(1)$  axially symmetry around the $z$-axis 
and the $L_z$ is a conserved quantity. Hence, the magnetic quantum number $m$ is  a good quantum number during the evolution process.  
In addition to the $U(1)$ symmetry, the Hamiltonian possesses the $\mathbb{Z}_2$ reflection symmetry with respect to the  $xy$-plane $(z=0)$: 
\be
z~~\rightarrow~~-z~~~~~\text{or}~~~~\theta~~\rightarrow~~\pi-\theta. 
\ee
As a result, the energy eigenvalues for $m$ and $-m$ are always degenerate. 

With the axial symmetry around the $z$-axis, we may assume that the eigenstates  of $H^{(\mu)}$ take  the form
\be
\psi^{(\mu)}(\theta, \phi)=\Theta^{(\mu)}(\theta) ~e^{im\phi}. 
\ee
The eigenvalue problem on the ellipsoid is therefore reduced to the 1D eigenvalue problem on the latitude with the polar angle  $\theta$: 
\be
-\frac{1}{2M}\biggl( \frac{1}{\rho^2}\frac{d^2}{d\theta^2} +\cot\theta \frac{1}{\rho^4}\frac{d}{d\theta}-\frac{1}{\sin^2\theta} (m+A_{\phi}(\theta))^2     \biggr)~\Theta(\theta) =E ~\Theta(\theta). \label{oneeqel} 
\ee
Solving (\ref{oneeqel}) in an analytical method  will be a formidable task. We then use the function of the $\it{Mathematica}$, ``NDEigensystem'', and  derive its eigenvalues and eigenstates  numerically.    
We will consider a deformation from a sphere  to an ellipsoid  by changing the parameter $\mu$. 

\begin{figure}[tbph]
\center
\includegraphics*[width=90mm]{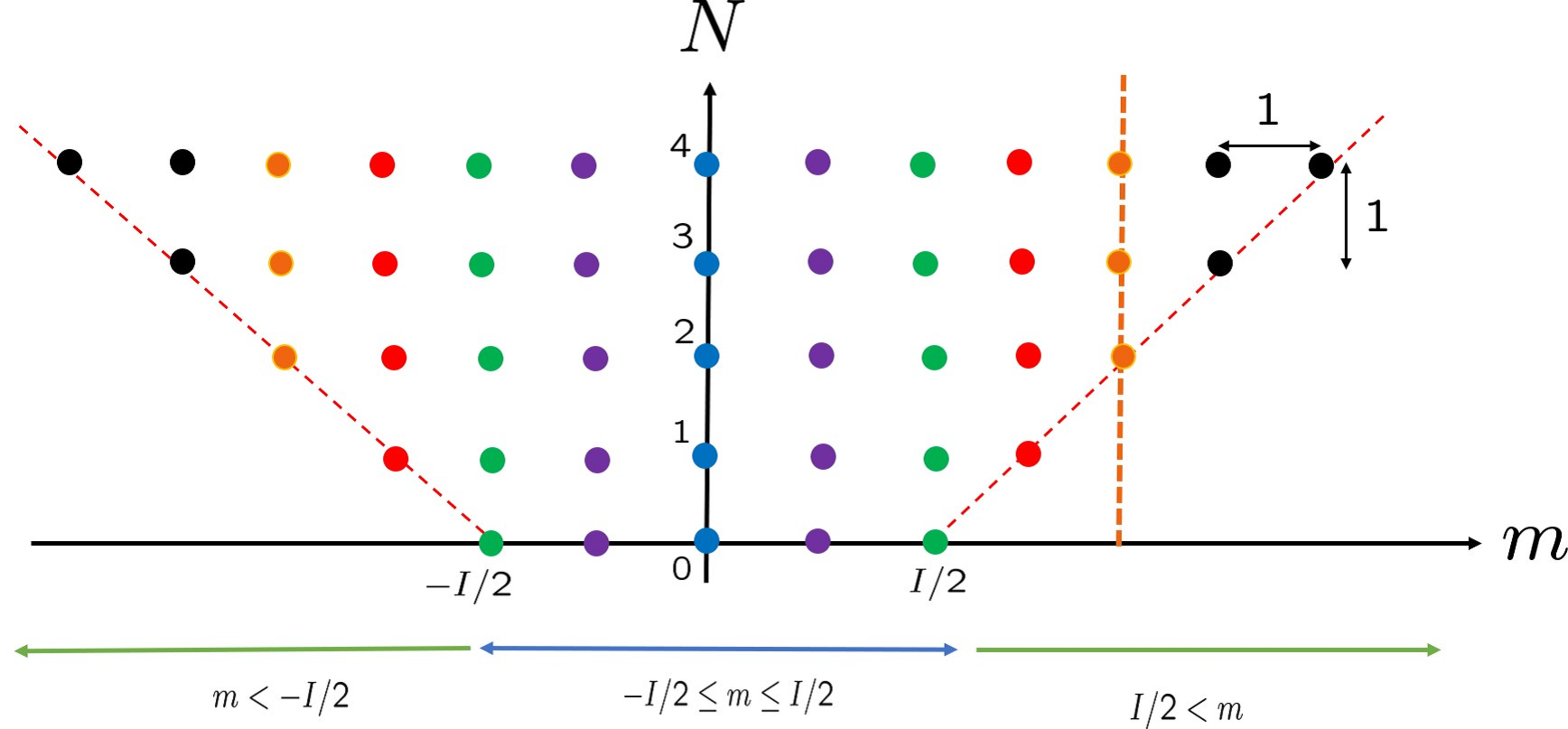}
\caption{The Landau levels with $I/2=2$ for the sphere ($\mu=1$) and the range of the magnetic quantum number. The vertical  dashed orange line  represents $m=\frac{I}{2}+2$, which concerns $N\ge 2$ Landau levels.}
\label{magm.fig}
\end{figure}

Before proceeding to  details, it may be worthwhile to explain how we can  solve (\ref{oneeqel}) numerically.  In the case of the sphere
 $(\mu=1)$, 
the Landau level is indexed by a non-negative integer $N$, and 
the range of the magnetic quantum number is given by  (\ref{rangem})    
(see Fig.\ref{magm.fig} which represents the allowed values of $m$ and $N$). 
 When we numerically solve the Landau problem  using (\ref{oneeqel}), 
 we first need to specify $m$ as an input parameter of (\ref{oneeqel}).  By solving (\ref{oneeqel}), we  later determine the energy eigenvalues of the Landau levels. 
From Fig.\ref{magm.fig}, we see that the allowed values of $N$ obviously depend on $m$ as 
\begin{subequations}
\begin{align}
&\text{For}~|m|\le \frac{I}{2},~~ ~~~ N=0,1,2,\cdots, \\
&\text{For}~|m| >  \frac{I}{2},~~ ~~~ N=|m|-\frac{I}{2},~|m|+1-\frac{I}{2},~|m|+2-\frac{I}{2},~|m|+3-\frac{I}{2},\cdots. \label{otherre}
\end{align}
\end{subequations}
Note that if  we enter $|m| > I/2$ in (\ref{oneeqel}), the  lowest energy obtained does $\it{not}$ start from the lowest Landau level $(N=0)$, but from the  $N=(|m|-\frac{I}{2})$th Landau level. 
 Similar care has  to be taken  when solving  (\ref{oneeqel}) numerically in the case of the ellipsoid: If we want to  know the evolution of the $N$th Landau level, we must first enter an $m$  $(|m|\le N+\frac{I}{2})$ in  Eq.(\ref{oneeqel})  and solve the eigenvalue equation numerically. Then we  choose the $N$th energy eigenvalue from the lowest for $|m|\le {I}/{2}$, while  $N-|m|+{I}/{2}$th for $|m|> I/2$. That eigenvalue signifies the evolved energy of the original $N$th Landau level eigenstate with quantum magnetic number $m$.  We repeat this procedure for all 
$m$s in the range, $-N-\frac{I}{2} \le m \le N+\frac{I}{2}$.

\subsection{Evolution of the Landau levels and  matrix geometries}

\begin{figure}[tbph]
\center
\includegraphics*[width=160mm]{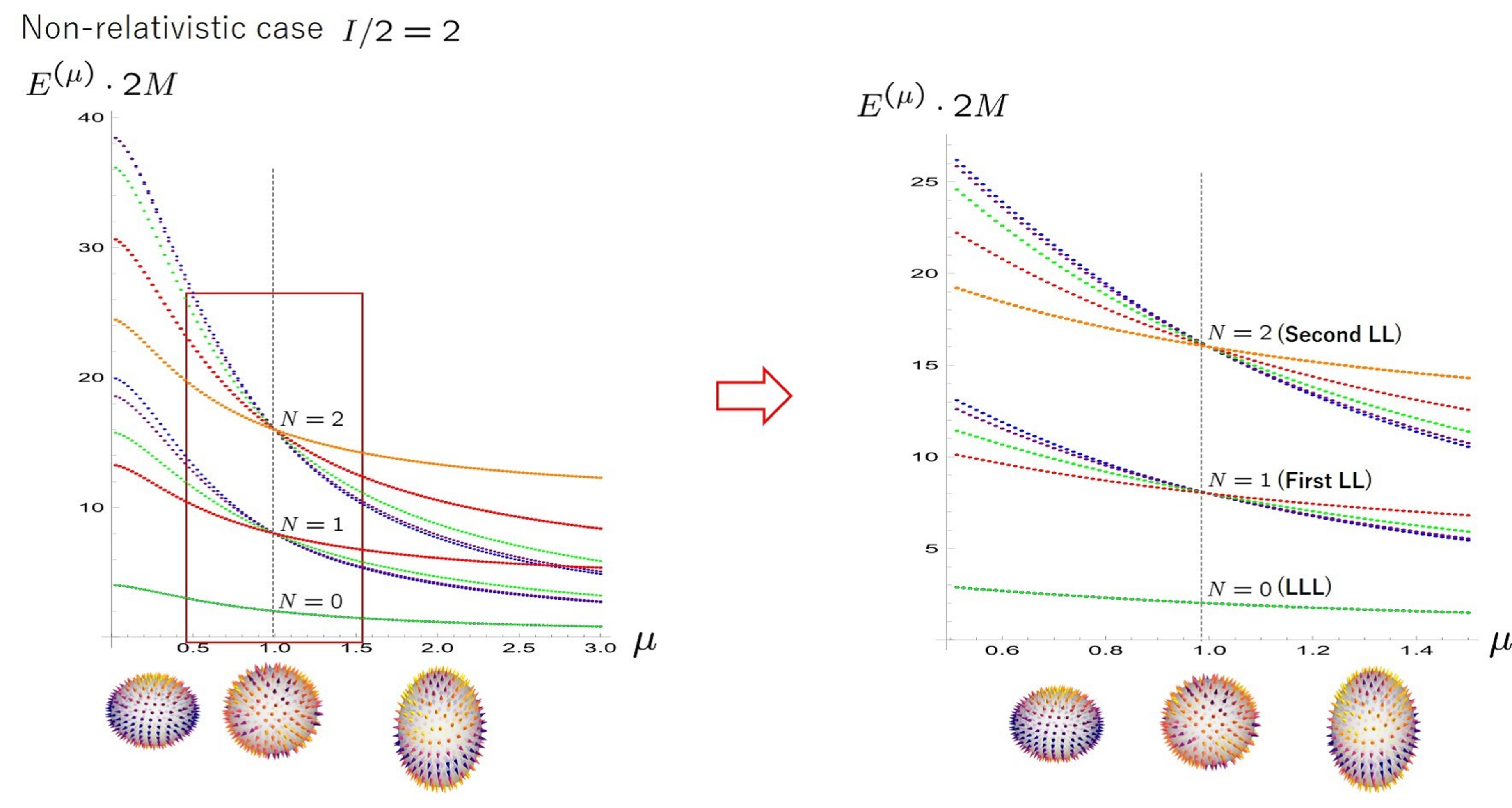}
\caption{The spectral flow of the non-relativistic Landau model for $I/2=2$. In the $N$th Landau level, there are $2N+5$ degenerate eigenstates $(-N-2\le m\le N+2)$.   
The blue, purple, green, red and orange curves correspond to $m=0$, $\pm 1$, $\pm 2$, $\pm 3$, and $\pm 4$, respectively (each of the curves is doubly degenerate except for the blue curve $m=0$). This coloring corresponds to  that of Fig.\ref{magm.fig}.  
At $\mu=1$, the energy levels converge to the Landau levels and the level crossings occur. Interestingly, the degeneracy of the lowest Landau level is not  lifted for arbitrary $\mu$.}
\label{nflow.fig}
\end{figure}

\begin{figure}[tbph]
\center
\includegraphics*[width=160mm]{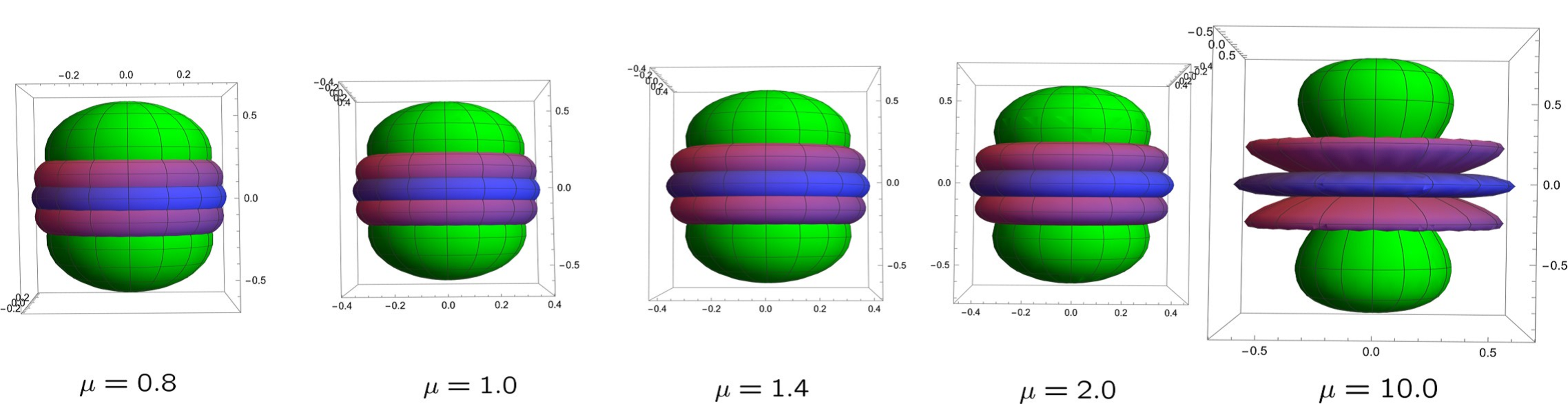}
\caption{The absolute values of the lowest Landau level eigenstates $\psi_{N=0, m}^{(\mu)}$ of $I/2=2$.  }
\label{lllwa.fig}
\end{figure}

We now solve the eigenvalue problem of the Landau Hamiltonian (\ref{landauham}) 
\be
H^{(\mu)}\psi^{(\mu)}_{N, m}(\theta, \phi) =E_{N, m}^{(\mu)}\psi_{N, m}^{(\mu)}(\theta, \phi),
\ee
where $\psi_{N, m}^{(\mu)}(\theta, \phi)$ are normalized on the ellipsoid: 
\be
\langle \psi^{(\mu)}_{N, m}|\psi^{(\mu)}_{N', m'}\rangle=\int_{\text{ellipsoid}} d\Omega~ \psi^{(\mu)}_{N, m}(\theta, \phi)^{*} ~ \psi^{(\mu)}_{N', m'}(\theta, \phi)=\delta_{N,N'}\delta_{m, m'}.
\ee
Here, $d\Omega$ is  the area element of the ellipsoid (\ref{areaelel}). In the sphere-limit ($\mu=1$),  the eigenstates are reduced to the monopole harmonics, $\psi_{N, m}^{(\mu=1)}(\theta, \phi)=Y_{N,m }(\theta, \phi)$.  Figure \ref{nflow.fig} represents the  spectral flow  of $E_{N,m}^{(\mu)}$. One may see that 
 the $N(\ge 1)$th Landau level is  split into $N+[I/2]+1$ energy levels ($[I/2]$ denotes the maximum integer that does not exceed $I/2$)\footnote{
For even $I$, each of the $N+[I/2]$ energy levels is doubly degenerate by the contribution of $\pm m$ and the  energy level with $m=0$ is non-degenerate. For odd $I$, each of the $N+[I/2]+1$ energy levels is doubly degenerate by the contribution of $\pm m$.  
} as $\mu$ changes from $\mu=1$. 
Figure \ref{nflow.fig} also shows that for the oblate spheroid  ($\mu<1$) the energy levels are in the inverse order of $|m|$, while for the prolate spheroid  ($\mu>1$) the energy levels are  in the order of $|m|$. 
The eigenstates are also derived numerically. 
From Fig.\ref{nflow.fig}, we find that the degeneracy of the lowest Landau level is not lifted, but the lowest Landau level eigenstates evolve in the degenerate level during the deformation.  
The absolute values of the evolving lowest Landau level eigenstates  are  depicted in Fig.\ref{lllwa.fig}. One may find that the set of the eigenvalues is generally elongated along the $z$ axis as $\mu$ increases.   This behavior qualitatively follows that of the classical ellipsoid. 

 We  apply the level projection  to the set of the   $N+[I/2]+1$ energy levels that originate from the $N$th Landau level: 
\be
(X_{i}^{(\mu)})_{mn} =\langle \psi^{(\mu)}_{N,m}|x_i|\psi^{(\mu)}_{N,n}\rangle=\int_{\text{ellipsoid}} d\Omega~ \psi^{(\mu)}_{N,m}(\theta, \phi)^{*} x_i \psi^{(\mu)}_{N,n}(\theta, \phi)
\ee
or 
\be
X_i^{(\mu)} =\int_{\text{ellipsoid}} d\Omega~ x_i(\psi_N^{(\mu)} {\psi_N^{(\mu)}}^{\dagger})^t . \label{xmatp}
\ee
The matrix $Z^{(\mu)}$ realizes the  $U(1)$ generator of the rotation around the $z$-axis. 
Since the Landau Hamiltonian (\ref{landauham}) respects only the $U(1)$ rotation symmetry,  the obtained matrix ellipsoids are expected to possess the $U(1)$ symmetry: 
\be
[{X^{(\mu)}}^2+{Y^{(\mu)}}^2 +\frac{1}{\mu^2} {Z^{(\mu)}}^2, ~Z^{(\mu)}] =0 \label{commux2z1}
\ee
or 
\be
[{X^{(\mu)}}^2+{Y^{(\mu)}}^2, ~Z^{(\mu)}] =0. \label{commux2z}
\ee
 Also recall that the $U(1)$ symmetry of the matrix geometry is anticipated from  the discussion in Sec.\ref{subsec:classellip}.  
  We  explicitly show (\ref{commux2z}) in the following general argument. 
Since $x\pm iy$ and $z$ carry the quantum magnetic numbers  $m=\pm 1$ and $0$, respectively, the selection rule tells 
\begin{subequations}
\begin{align}
&(X^{(\mu)}+iY^{(\mu)})_{mn} := \langle \psi^{(\mu)}_{N,m}|(x+iy)|\psi^{(\mu)}_{N,n} \rangle=g^{(\mu)}_m\delta_{m,n+1}, 
\\
&(X^{(\mu)}-iY^{(\mu)})_{mn}:= \langle \psi^{(\mu)}_{N,m}|(x-iy)|\psi^{(\mu)}_{N,n}\rangle={g^{(\mu)}_n}^* \delta_{m+1,n}, 
\\
&(Z^{(\mu)})_{\alpha\beta}:= \langle \psi^{(\mu)}_{N,m}|z|\psi^{(\mu)}_{N,n}\rangle =f^{(\mu)}_m\delta_{mn}, 
\end{align}
\end{subequations}
with $f^{(\mu)}_m \in \mathbb{R}$ and $g^{(\mu)}_m \in \mathbb{C}$. The mirror symmetry about the $xy$ plane implies that  $|g^{(\mu)}_{-m}|=|g^{(\mu)}_{m+1}|$ and $f^{(\mu)}_{-m}=-f^{(\mu)}_m$.  
We therefore have 
\begin{align}
({X^{(\mu)}}^2 +{Y^{(\mu)}}^2)_{mn} &=\frac{1}{2}((X^{(\mu)}+iY^{(\mu)})(X^{(\mu)}-iY^{(\mu)}) +(X^{(\mu)}-iY^{(\mu)})(X^{(\mu)}+iY^{(\mu)}))_{mn}\nn\\
&=\frac{1}{2}(|g^{(\mu)}_m|^2+|g^{(\mu)}_{m+1}|^2) \delta_{mn}=({X^{(\mu)}}^2 +{Y^{(\mu)}}^2)_{-m,-n}. \label{xsqysq}
\end{align}
Both ${X^{(\mu)}}^2+{Y^{(\mu)}}^2$ and $Z^{(\mu)}$ are  diagonal matrices, and so they obviously satisfy  (\ref{commux2z}). 
Using the numerically derived matrix coordinates, we can also confirm that the matrix geometries  do not have the $SU(2)$ symmetries generated by $X^{(\mu)}$ and $Y^{(\mu)}$:  
\be
[{X^{(\mu)}}^2+{Y^{(\mu)}}^2  +\frac{1}{\mu^2} {Z^{(\mu)}}^2, ~X^{(\mu)}] \neq 0, ~~~~~[{X^{(\mu)}}^2+{Y^{(\mu)}}^2+ \frac{1}{\mu^2} {Z^{(\mu)}}^2, ~Y^{(\mu)}] \neq 0.
\ee
As $Z^{(\mu)}$ is a diagonal matrix, 
its diagonal matrix components $f_m^{(\mu)}$  do not depend on the phases of the basis states. Meanwhile,  
 $X^{(\mu)}$ and $Y^{(\mu)}$ have off-diagonal components $g_m$ that  depend on the relative phases of the basis states.   However, as indicated by (\ref{xsqysq}),   the sum of their squares is determined only by the absolute values of $g$s. Therefore,  the matrix counterpart  of (\ref{defellipsoid}) 
\be
{X^{(\mu)}}^2+{Y^{(\mu)}}^2+\frac{1}{\mu^2}{Z^{(\mu)}}^2
\ee
is not relevant to the phases of the basis states.

\begin{figure}[tbph]
\center
\includegraphics*[width=160mm]{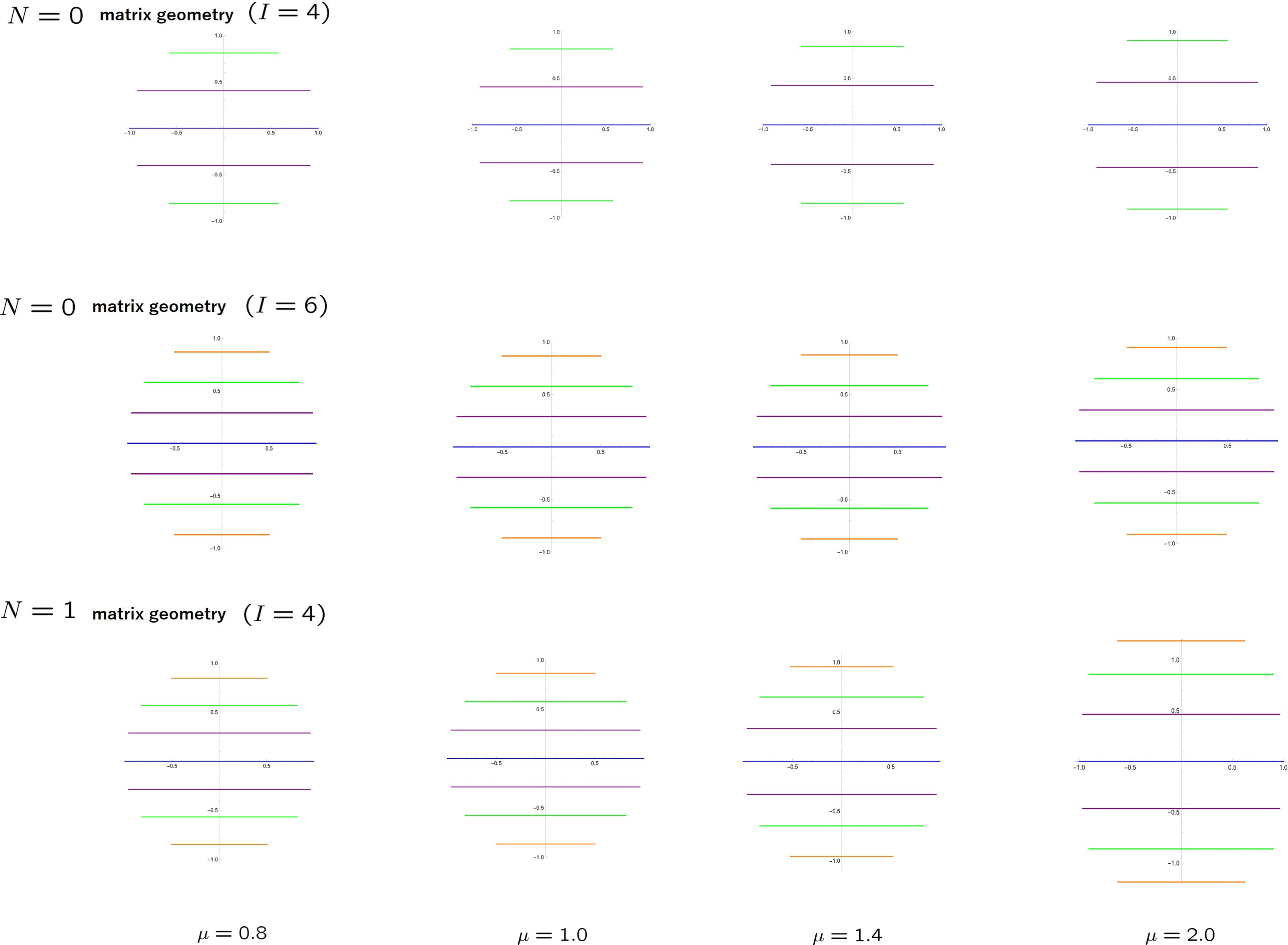}
\caption{Ellipsoidal matrix geometries in the Landau levels. At $\mu=1$, the fuzzy sphere geometry is reproduced. The horizontal lengths of the latitudes of the ellipsoidal matrix geometries represent the eigenvalues of $2\sqrt{(\hat{X}^{(\mu)})^2+(\hat{Y}^{(\mu)})^2}$.  
The second and third rows denote the evolutions of the matrix geometries  with  $7\times 7$ matrix size of the lowest Landau level and the first Landau level, respectively: One may find that the behavior of the  ellipsoidal matrix geometry of the lowest Landau level geometry ($N=0$)  differs from those of the first Landau level ($N=1$).  
}
\label{nonrelfu.fig}
\end{figure}

As discussed in Sec.\ref{subsec:matfuzzsphe},   $\it{each}$ Landau level develops its own matrix geometry.
Figure \ref{nonrelfu.fig} exhibits the numerically obtained ellipsoidal matrix geometries of the lowest Landau level  ($N=0$) and the first Landau level ($N=1$). 
To reduce the overall scaling, we adopt the normalized matrix coordinates: 
\be
\hat{X}_i^{(\mu)}:= \frac{1}{\text{max}(\sqrt{{X^{(\mu)}}^2+{Y^{(\mu)}}^2})} X^{(\mu)}_i, \label{normamatco}
\ee
so that $\text{max}((\hat{X}^{(\mu)})^2+(\hat{Y}^{(\mu)})^2)=1$. 
It can be seen that the behaviors of the ellipsoidal matrix geometries are quantitatively different in each  Landau level, but qualitatively similar:  As $\mu$ increases,  
the matrix geometries are generally elongated along the $z$ axis analogous to  the eigenstates  (Fig.\ref{lllwa.fig}) (for $\mu <1$, this tendency is not  clear).

\section{Relativistic Landau model on an ellipsoid}\label{sec:rella}

In this section, we generalize the previous non-relativistic analysis to the relativistic case.  We consider a relativistic particle on an ellipsoid under the magnetic field.

\subsection{Dirac-Landau operator }

We introduce gamma matrices, zweibein, and the spin-connection  as (see Appendix \ref{append:rieman} for details)
\begin{align}
&\gamma^{m=1,2}=\{\sigma_1, \sigma_2\}, ~~~e_1^{~~\theta}=\frac{1}{\rho},~~e_2^{~~\phi} =\frac{1}{\sin\theta}, ~~~\text{other}~e_m^{~~\mu}~~\text{are zeros}, \nn\\
&\omega_{\theta} =0, ~~~~\omega_{\phi}(\theta) =\omega_{\phi 12}(\theta)\sigma_{12} =-\frac{1}{2\rho}\cos\theta  \sigma_3. 
\end{align}
With these, we construct the Dirac-Landau operator on the ellipsoid: 
\be
-i\fsl{\mathcal{D}} =-i\sum_{m=1,2}\sum_{\mu=\theta, \phi}e_m^{~~\mu}\gamma^m \mathcal{D}_{\mu} . \label{covderivs}
\ee
Here, the covariant derivatives contain both  the spin-connection and the gauge connection, 
\be
\mathcal{D}_{\mu} := \frac{\partial}{\partial x^{\mu}}+i(\omega_{\mu}\otimes {1}-\bs{1}_2\otimes A_{\mu}),
\ee
where the  gauge field is  the same as (\ref{gaugefiele})
\be
A_{\theta}=0, ~~~~A_{\phi} =\frac{2\pi I}{S}\Omega_{\phi}(\theta). \label{basiccomore}
\ee
The covariant derivatives (\ref{covderivs}) are  expressed as 
\be
\mathcal{D}_{\theta} =\partial_{\theta}, ~~~~~~\mathcal{D}_{\phi} =\partial_{\phi} +i\omega_{\phi}-iA_{\phi}=\partial_{\phi}-i\frac{1}{2}\frac{\cos\theta}{\rho} \sigma_3-i\frac{2\pi I}{S}\Omega_{\phi} , 
\label{caldcov}
\ee
and then 
\begin{align}
-i\fsl{\mathcal{D}} =-ie_1^{~\theta}\sigma_1\frac{\partial}{\partial\theta} -ie_2^{~\phi}\sigma_2 (\frac{\partial}{\partial\phi} +i(\omega_\phi-A_{\phi})). \label{diracoprel}
\end{align}
The Dirac-Landau operator respects the chiral symmetry for arbitrary values of $\mu$, 
\be
\{\sigma_3, -i\fsl{\mathcal{D}}\}=0.
\ee
Hence, the spectra of the Dirac-Landau operator are symmetric with respect to the zero energy. The Dirac-Landau operator is  commutative with  $L_z=-i\frac{\partial}{\partial\phi}$: 
\be
[L_{z} ,  -i\fsl{\mathcal{D}}]=0, \label{lzdfsl}
\ee
which implies that the present system is axially symmetric around the $z$-axis, and the magnetic quantum number can be used as a good quantum number, same as in the non-relativistic case. 
In the matrix representation, the Dirac-Landau operator (\ref{diracoprel}) is expressed as 
\be
-i\fsl{\mathcal{D}}=-i\frac{1}{\rho} \sigma_1(\frac{\partial}{\partial\theta} +\frac{1}{2}\cot\theta) -i\frac{1}{\sin\theta}\sigma_2(\frac{\partial}{\partial\phi} -i\frac{2\pi I}{S}\Omega_{\phi})=-i 
\begin{pmatrix}
0 & \eth_-\\
\eth_+ & 0 
\end{pmatrix}, \label{diraclandaumatel}
\ee
where 
\be
\eth_\pm := \frac{1}{\rho} \frac{\partial}{\partial\theta} +\frac{1}{2\rho}\cot\theta \pm i\frac{1}{\sin\theta}(\frac{\partial}{\partial\phi}-i\frac{2\pi I}{S}\Omega_{\phi}). \label{defedth}
\ee 

\begin{figure}[tbph]
\center
\includegraphics*[width=110mm]{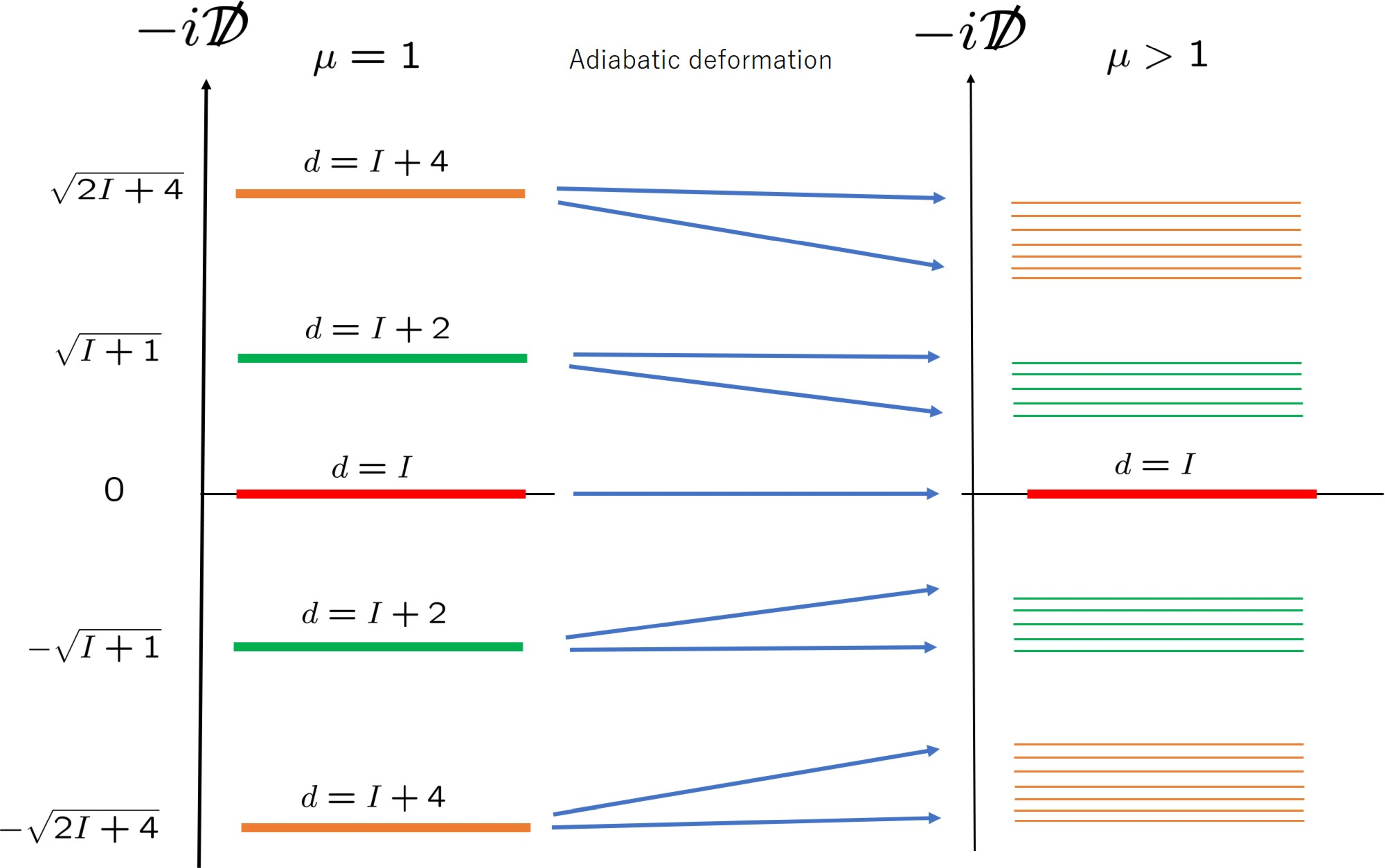}
\caption{The evolution of the spectra of the Dirac-Landau operator. The spectra of the Dirac-Landau operator are symmetric with respect to the zeroth Landau level. For $\mu=1$, the relativistic Landau level has the degeneracy due to the $SU(2)$ rotational symmetry.  The degeneracy of the zeroth Landau level is not lifted under the deformation as guaranteed by the Atiyah-Singer index theorem 
(see Sec.\ref{subsec:zeromode}).  }
\label{splitrel.fig}
\end{figure}

At the reference point $(\mu=1)$, (\ref{diraclandaumatel}) is reduced to  
\be
-i\fsl{\mathcal{D}}|_{\mu=1} = -i\begin{pmatrix}
0 & \frac{\partial}{\partial \theta} -i\frac{1}{\sin\theta}\frac{\partial}{\partial \phi}+(\frac{I}{2}+\frac{1}{2})\cot\theta \\
\frac{\partial}{\partial \theta} +i\frac{1}{\sin\theta}\frac{\partial}{\partial \phi}-(\frac{I}{2}-\frac{1}{2})\cot\theta & 0 
\end{pmatrix}, \label{diracopmu1}
\ee
which respects the $SU(2)$ rotational symmetry \cite{Hasebe-2016}.
The eigenvalues  are given by 
\be
-i\fsl{\mathcal{D}}|_{\mu=1}=\pm   \sqrt{N(N+I)} ~~~~~~~(N=0, 1, 2,   \cdots) \label{dreig}
\ee
with $I+2N$ degeneracy coming from the $SU(2)$ rotational symmetry:   
\be
m=\overbrace{N+\frac{I}{2}-\frac{1}{2}, ~N+\frac{I}{2}-\frac{3}{2}, ~\cdots, ~-(N+\frac{I}{2}-\frac{1}{2})}^{I+2N}. 
\label{relaqmagm}
\ee
The difference to the non-relativistic case (\ref{rangem}) is the 1/2 reduction of the $SU(2)$ angular momentum index.  
For arbitrary $\mu$,  the Dirac-Landau operator only has  the axial symmetry around the $z$-axis, and then the $SU(2)$ degeneracy of the relativistic Landau level (\ref{dreig}) is  lifted (Fig.\ref{splitrel.fig}).

\subsection{Supersymmetric Landau model }

 A standard strategy  for deriving the spectra of the Dirac-Landau operator is to take a square of the Dirac-Landau operator and transform it into a diagonal matrix with diagonal components of differential operators.    
 
From a generalized Lichnerowicz formula \cite{Dolan2003}, the square of the Dirac-Landau operator is given by  
\be
(-i\fsl{\mathcal{D}})^2 =-\Delta -e^{~~\mu}_{m}e^{~~\nu}_{n}\sigma_{mn} \otimes F_{\mu\nu}+\frac{\mathcal{R}}{4}.
\ee
In the present case, we have 
\be
(-i\fsl{\mathcal{D}})^2 =-\Delta -e^{~~\theta}_{1}e^{~~\phi}_{2}\sigma_{3}  F_{\theta\phi}  +\frac{\mathcal{R}}{4} =-\Delta -\frac{1}{\rho~\sin\theta}\sigma_{3}  F_{\theta\phi}  +\frac{\mathcal{R}}{4}, \label{dlsq}
\ee
where $\mathcal{R}:= 2\frac{\mu^2}{\rho^4}$ represents the scalar curvature and the Laplace-Beltrami operator is 
\be
\Delta := \frac{1}{\sqrt{g}} \mathcal{D}_{\mu} (\sqrt{g}g^{\mu\nu}\mathcal{D}_{\nu})= \frac{1}{\rho^2 }\partial^2_{\theta}+\cot\theta \frac{1}{\rho^4}\partial_{\theta} +\frac{1}{\sin^2\theta}\biggl(\partial_{\phi}-i\frac{1}{2}\frac{\cos\theta}{\rho} \sigma_3-i\frac{2\pi I}{S}\Omega_{\phi}\biggr)^2. 
\ee
For the field strength (\ref{prefield}), (\ref{dlsq}) is expressed as\footnote{One can also explicitly derive (\ref{sqdex}) using (\ref{diraclandaumatel}).}   
\be
H_{\text{susy}}:= (-i\fsl{\mathcal{D}})^2 
=-\Delta +\frac{1}{2}\frac{\mu^2}{\rho^4} -\frac{2\pi I}{S}\sigma_3
=\begin{pmatrix}
 H_+ & 0 \\
 0 & H_-
 \end{pmatrix}, 
\label{sqdex}
\ee
where 
\be
H_\pm := -\eth_\mp \eth_\pm =-\frac{1}{\rho^2 }\partial^2_{\theta}-\cot\theta \frac{1}{\rho^4}\partial_{\theta} -\frac{1}{\sin^2\theta}\biggl(\partial_{\phi}\mp i\frac{1}{2}\frac{\cos\theta}{\rho} -i\frac{2\pi I}{S}\Omega_{\phi}\biggr)^2 +\frac{1}{2}\frac{\mu^2}{\rho^4} \mp \frac{2\pi I}{S}. 
\ee
Here, $\eth_\pm$ are given by (\ref{defedth}). 
 We analyze the differential operators, $H_+$ and $H_-$, which appear as  the diagonal components of  (\ref{sqdex}).     
The operators $H_+$ and $H_-$ are exchanged under $\phi \rightarrow -\phi$ and $I \rightarrow -I$ and  exhibit the same energy spectra except for the zero energy: For $I>0$ $(I<0)$, $H_{+}$ $(H_-)$ accommodates the zero-modes. 

Supersymmetry is  manifest when we rewrite (\ref{sqdex}) as 
\be
H_{\text{susy}} =\{Q, \bar{Q}\}, 
\ee
with supercharges  
\be
Q=\fsl{D}_+:= \frac{1}{2}(1+ \sigma_3)\fsl{D}= \begin{pmatrix}
 0 & \eth_- \\
 0 & 0
 \end{pmatrix},~~~~~~\bar{Q}=-\fsl{D}_-:= -\frac{1}{2}(1- \sigma_3)\fsl{D}= -\begin{pmatrix}
 0 & 0 \\
 \eth_+ & 0
 \end{pmatrix},
\ee
which are obviously nilpotent operators, $Q^2=\bar{Q}^2=0$. It is easy to see that (\ref{sqdex}) is invariant under the  supersymmetric transformations, $[Q, H_{\text{susy}} ]=[\bar{Q}, H_{\text{susy}}]=0$, and then  the  $H_{\text{susy}}$ is  a supersymmetric  Hamiltonian on an ellipsoid. The $H_{\text{susy}}$ is commutative with the chirality matrix, $[H_{\text{susy}}, \sigma_3]=0$, while the supercharges are  anti-commutative with the chirality matrix, $\{Q, \sigma_3\}=\{\bar{Q},\sigma_3\}=0$. 
Therefore, there are degenerate energy eigenstates of $H_{\text{susy}}$ with opposite chirality, which are exchanged by the supersymmetric transformations:
\be
|E, \sigma_3=+1\rangle ~~~\overset{\bar{Q}}{\longrightarrow}~~~|E, \sigma_3=-1\rangle ~~\overset{{Q}}{\longrightarrow}~~~|E, \sigma_3=+1\rangle ~~~(E\neq 0 ).
\ee
The degenerate eigenstates with opposite chiralities realize the supersymmetric partners.  
The supersymmetric transformation vanishes the zero-modes: 
\be
Q|E=0,\sigma_3\rangle=\bar{Q}|E=0,\sigma_3\rangle =0, 
\ee
which indicates that the zero-modes are the supersymmetric invariant states. The zero-modes can be taken  as the eigenstates of $\sigma_3$ and  satisfy   
\be
-i\fsl{\mathcal{D}}\begin{pmatrix}
\psi_m \\
0\end{pmatrix}=0~~\rightarrow~~(-i\fsl{\mathcal{D}})^2\begin{pmatrix}
\psi_{m} \\
0\end{pmatrix}=H_{\text{susy}}\begin{pmatrix}
\psi_m \\
0\end{pmatrix} =0.  
\ee
The zeroth Landau level eigenstates are equal to the zero-mode eigenstates of the Dirac-Landau operator. Notice that the above discussion holds for arbitrary $\mu$ and that  supersymmetry is preserved for the  deformation.

 At the reference point $\mu=1$, $H_{\text{susy}}$ (\ref{sqdex}) is reduced to the supersymmetric Landau Hamiltonian on a sphere \cite{Hasebe-2016}: 
\be
H_{\text{susy}}|_{\mu=1}= -\partial^2_{\theta}-\cot\theta \partial_{\theta} -\frac{1}{\sin^2\theta}\biggl(\partial_{\phi}+i\frac{1}{2}(I-1)\cos\theta\biggr)^2 +\frac{1}{2}-\frac{ I}{2}\sigma_3,
\ee
with eigenvalues 
\begin{align}
N(N+I) ~~~(N=0,1,2,\cdots)
\end{align}
and degeneracies  
\be
d_{N=0}=I+2N, ~~~~d_{N=1,2,3,\cdots}=2(I+2N).  
\ee

\begin{figure}[tbph]
\center
\includegraphics*[width=140mm]{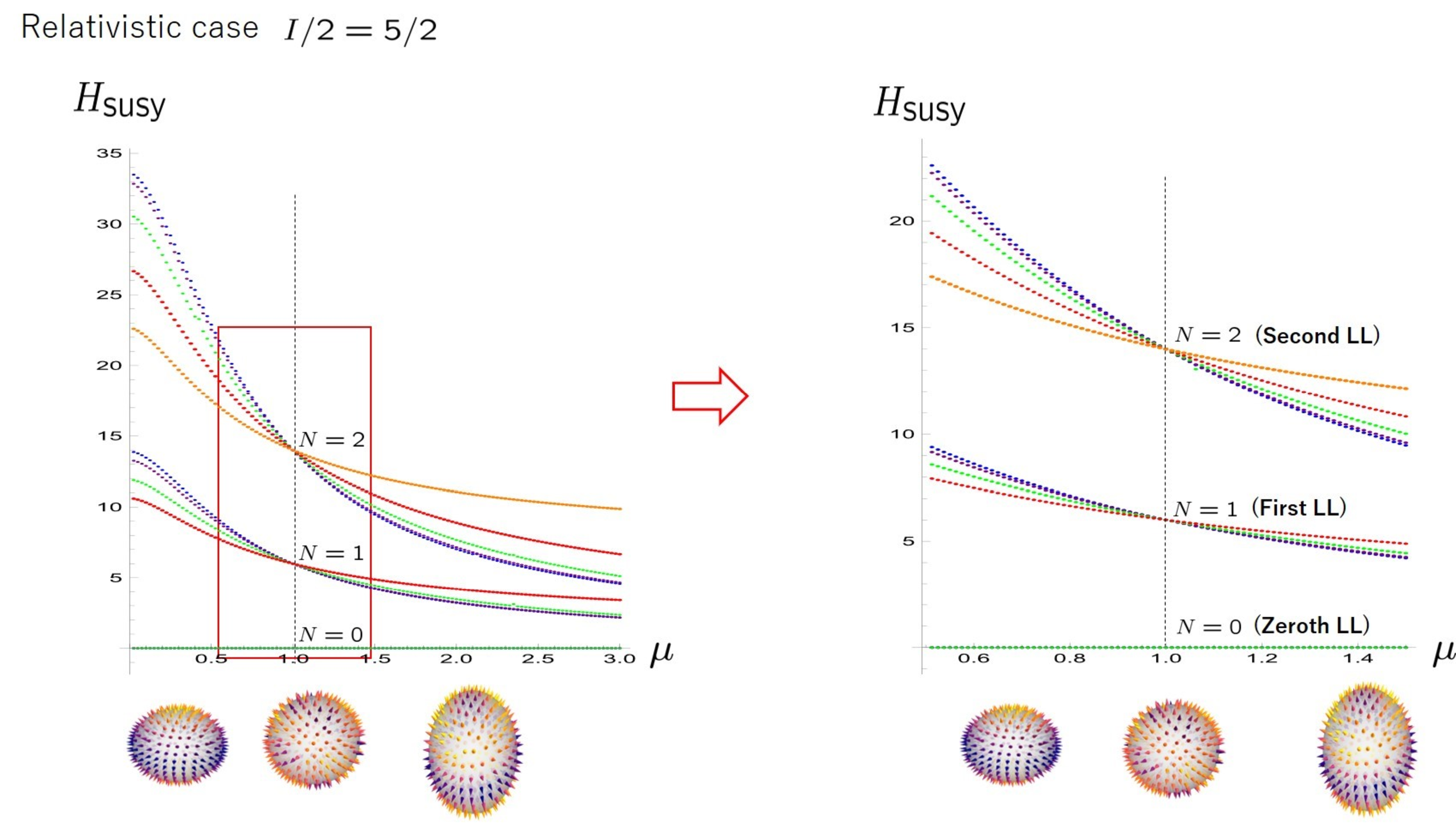}
\caption{The spectral flow of the supersymmetric Landau model (\ref{sqdex}) for $I/2=5/2$.  Every spectrum monotonically decreases   as $\mu$ increases. Note  that the degeneracy of the zeroth Landau level  is not lifted for any $\mu$.   
The $N(\ge 1)$th Landau level accommodates $2\times (2N+I)=4N+10$  degeneracy. (Here, $2\times $ comes from the spinor degrees of freedom.) The zeroth  Landau level accommodates   $I=5$ degeneracy.   
The blue, purple, green, red and orange curves correspond to $m=0$, $\pm 1$, $\pm 2$, $\pm 3$, and $\pm 4$, respectively. Because of the mirror symmetry about the $xy$-plane, the $m$ and $-m$ eigenstates are always degenerate.  
}
\label{rflow.fig}
\end{figure}

\subsection{Zero-modes and the Atiyah-Singer index theorem}\label{subsec:zeromode}

We discuss the zero-modes of the Dirac-Landau operator on an ellipsoid and their relation to the Atiyah-Singer index theorem. Let us begin with the reference point $\mu=1$, at which the spherical relativistic Landau model on the two-sphere is realized.  
The zero-modes can be taken  as the eigenstates of $\sigma_3$ and  satisfy   
\be
-i\fsl{\mathcal{D}}\begin{pmatrix}
\psi_m \\
0\end{pmatrix}=0~~\rightarrow~~-i\fsl{\mathcal{D}}_-\begin{pmatrix}
\psi_m \\
0\end{pmatrix}=0,~
\ee
where the zero-modes $\psi_m$ $(m=\overbrace{-\frac{I-1}{2}, -\frac{I-1}{2}+1, \cdots, \frac{I-1}{2})}^{I}$ ($N=0$ in (\ref{relaqmagm})) are given by the monopole harmonics at $\mu=1$ \cite{Hasebe-2016}. The index of the Dirac-Landau operator on $S^2$ is then obtained as  
\be
\text{Ind}(-i\fsl{D}) := \dim \text{Ker}(-i\fsl{D}_-) -\dim \text{Ker}(-i\fsl{D}_+) =I-0=I. 
\ee
Meanwhile, the number of the magnetic fluxes is given by 
\be
\frac{1}{2\pi}\int_{S^2} F=I, 
\ee
where $F=\frac{I}{2}\sin\theta d\theta \wedge d\phi$.
This exemplifies the  Atiyah-Singer index  theorem on $S^2$: 
\be
\text{Ind}(-i\fsl{\mathcal{D}})|_{S^2} =\frac{1}{2\pi}\int_{S^2} F. \label{index2d}
\ee
 The index theorem  should hold under a continuous deformation of base-manifold. 
 The right-hand side of (\ref{index2d}) denotes the number of the fluxes penetrating the surface enclosing the monopole, which is invariant under the continuous deformation of the surface due to  Gauss's theorem.\footnote{To be precise, the present gauge field also evolves in the deformation process, but still (\ref{ffequiv}) holds.  See (\ref{1stchern}).  }  We therefore have   
\be
I=\frac{1}{2\pi}\int_{S^2} F=\frac{1}{2\pi}\int_{\text{ellipsoid}} F. \label{ffequiv}
\ee
Meanwhile, it is not so obvious that the index of the Dirac-Landau operator does not change under the continuous deformation, because the  energy levels generally evolve and the degeneracies of energy levels are usually lifted under the deformation.  
For the zeroth Landau level, however, this is not the case:   Figure \ref{rflow.fig}  clearly shows that the zeroth Landau level remains at the zero energy and the degeneracy is not lifted during  deformation.  
Thus,  the Dirac-Landau operator  on the ellipsoid  always accommodates the same number of zero-modes as that on the two-sphere:        
\be
\text{Ind}(-i\fsl{\mathcal{D}})|_{\text{ellipsoid}}= \text{Ind}(-i\fsl{\mathcal{D}})|_{S^2}=I. 
\ee
Consequently  we  confirmed 
\be
\text{Ind}(-i\fsl{\mathcal{D}})|_{\text{ellipsoid}}= I=\frac{1}{2\pi}\int_{\text{ellipsoid}} F, 
\ee
which demonstrates the Atiyah-Singer index theorem.

\section{Ellipsoidal matrix geometry and fuzzy ellipsoid }\label{sec:diffmatelfuz}

We investigate basic properties of the  ellipsoidal matrix geometries. 
\subsection{Ellipsoidal matrix geometries}\label{subsec:ellipmatrel}

\begin{figure}[tbph]
\center
\includegraphics*[width=160mm]{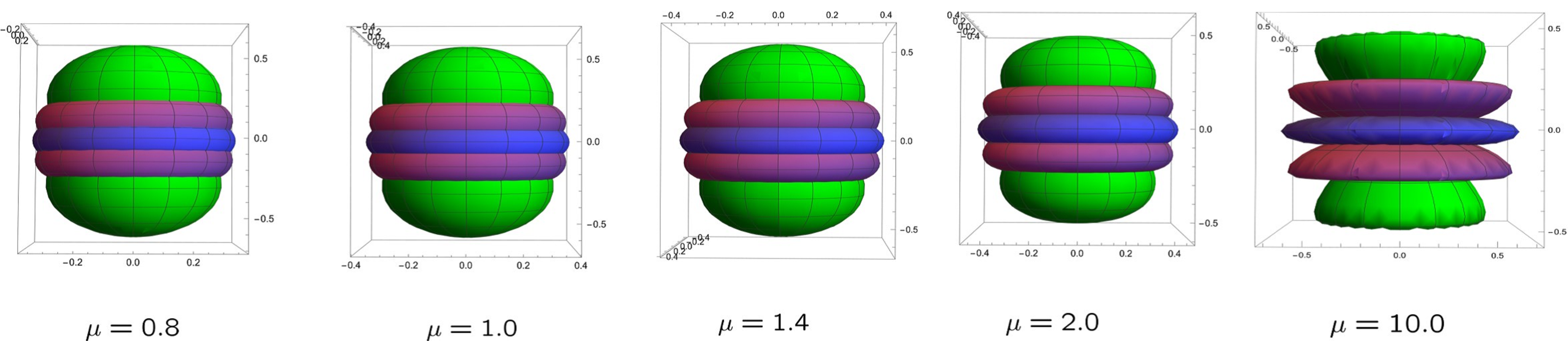}
\caption{Absolute values of the zero-modes of $I/2=5/2$.  }
\label{zewa.fig}
\end{figure}

\begin{figure}[tbph]
\center
\includegraphics*[width=140mm]{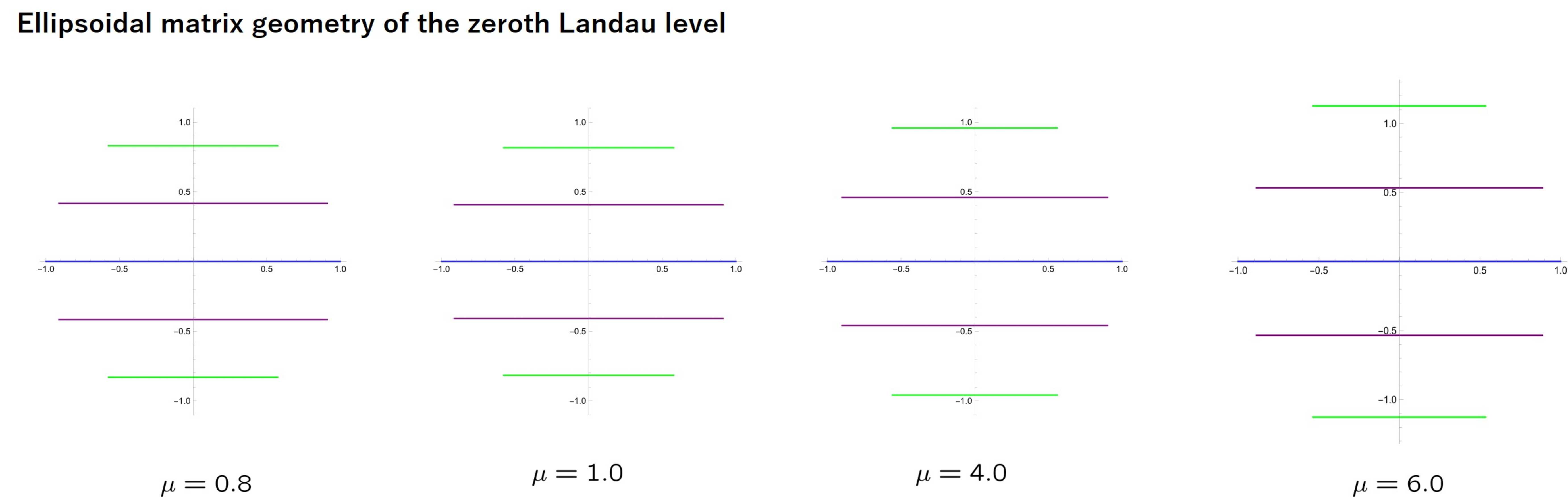}
\caption{Evolution of the ellipsoidal matrix geometry in the zeroth Landau level of $I/2=5/2$. At $\mu=1$, the fuzzy sphere geometry is realized. The ellipsoidal matrix geometry shows  qualitatively similar behavior to the deformation of the  original classical ellipsoid. }
\label{relafuzz.fig}  
\end{figure}

\begin{figure}[tbph]
\center
\includegraphics*[width=145mm]{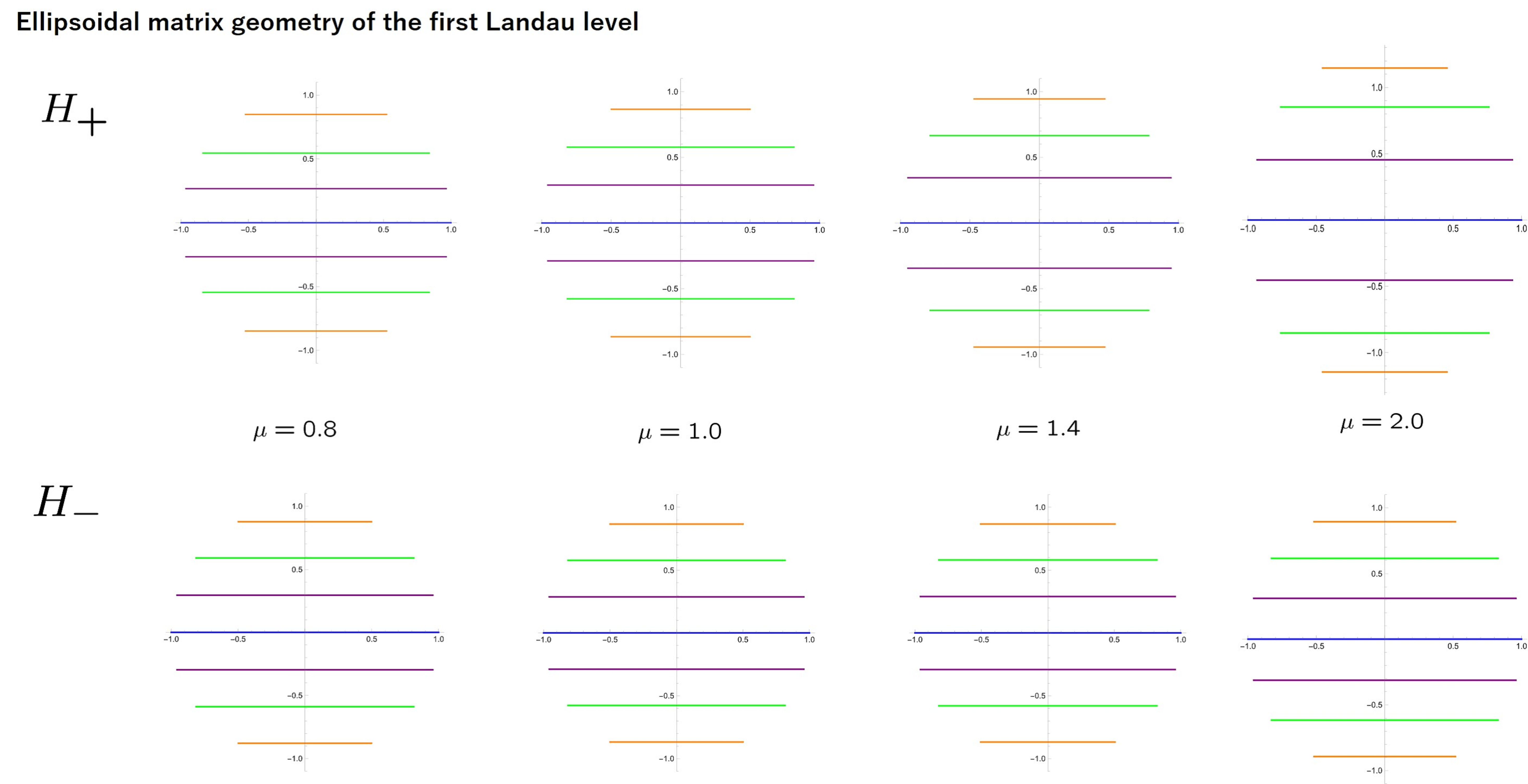}
\caption{Evolution of the ellipsoidal matrix geometries  in the supersymmetric first Landau level of $I/2=5/2$.  The upper/lower ellipsoidal matrix geometries are derived from the first Landau level eigenstates of $H_{+/-}$.   }
\label{N1rel.fig} 
\end{figure}

We discuss the ellipsoidal matrix geometry of the relativistic Landau model. Figure \ref{zewa.fig} shows the behaviors of the zero-mode wave-functions obtained by the numerical calculations. The distribution of the eigenstates is generally elongated along the $z$ direction as the deformation parameter increases.    
Figures \ref{relafuzz.fig} and \ref{N1rel.fig} exhibit the matrix ellipsoids of the zeroth Landau level and the first Landau level of the supersymmetric Landau model.  The energy levels of  $H_+$ and $H_-$ are  identical (except for the zero-energy), but the corresponding matrix geometries are different as shown in Fig.\ref{N1rel.fig}.


\begin{figure}[tbph]
\center
\hspace{-0.4cm}
\includegraphics*[width=160mm]{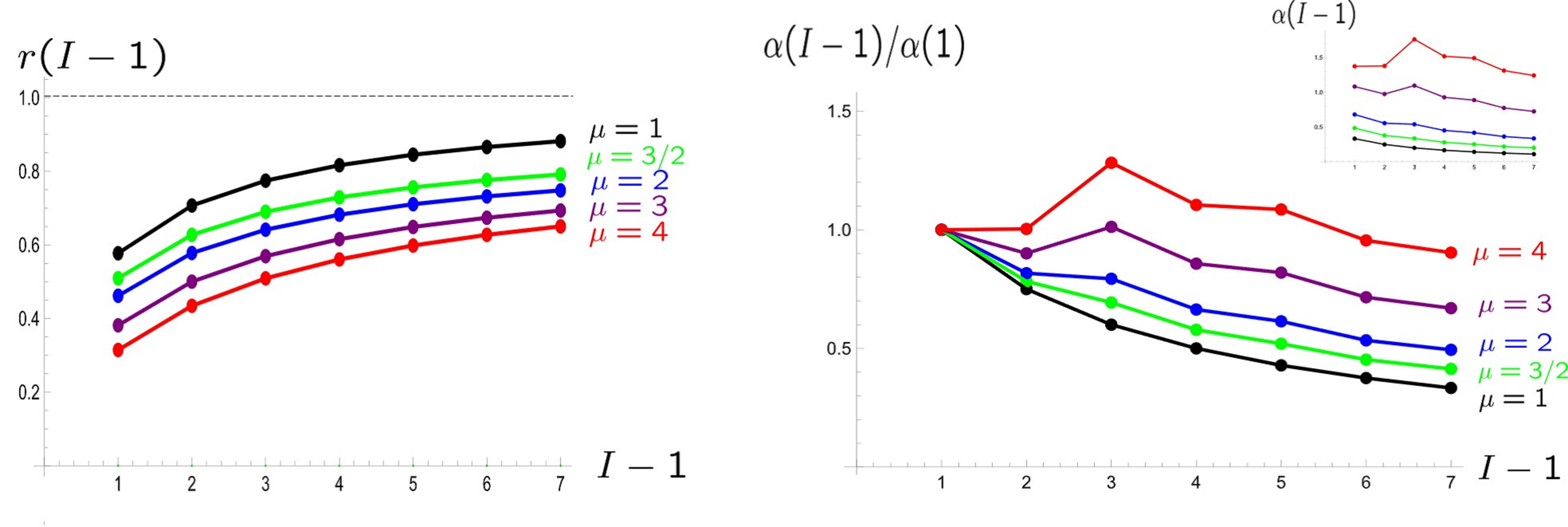}
\caption{Left: the behaviors of (\ref{rradform}).    Right:  the ratio $\alpha(I-1)/\alpha(1)$ (\ref{ncsellip}).
 }
\label{rrad.fig} 
\end{figure}

In the following, we discuss the commutative limit $(I~\rightarrow~\infty)$ of the ellipsoidal matrix geometry in the zeroth Landau level. 
We introduce the ``radius'' as 
\be
r(I-1):= \frac{1}{I}\tr\biggl(\sqrt{{X}^2+{Y}^2+\frac{1}{\mu^2}{Z}^2}\biggr), \label{rradform}
\ee
and,  considering the first equation of (\ref{poisellip}),  we also introduce the ``non-commutative scale'' as
\be
\alpha(I-1):= \sum_{m>0}^{(I-1)/2}\frac{([X, Y])_{mm}}{\biggl( i \frac{1}{\sqrt{\mu^4 (x^2+y^2)+z^2}}z\biggr)_{mm}}. \label{ncsellip}  
\ee
In the sphere limit $\mu\rightarrow 1$,  (\ref{rradform}) and  (\ref{ncsellip}) are reduced to the radius (\ref{zeromoderad}) and the non-commutative scale (\ref{zeromodexcoord}) of the fuzzy sphere.\footnote{
(\ref{ncsellip}) is reduced to (\ref{zeromodexcoord}) as 
\be
\alpha(I-1)|_{\mu=1}= i\sum_{m>0}^{(I-1)/2}\frac{([X, Y])_{mm}}{( Z)_{mm}}=\frac{2}{I+1}.  \label{ratisph}
\ee
}    
Figure \ref{rrad.fig} depicts the behaviors of (\ref{rradform}) (left) and the ratios $\alpha(I-1)/\alpha(1)$  (right).
One may find  that  $\lim_{I\rightarrow \infty}r(I-1)=1$ and the  ratios $\alpha(I-1)/\alpha(1)$ decrease as  $I$ increases, which  imply that the  ellipsoidal matrix geometries   approach the commutative ellipsoids with unit radius.   

\subsection{Comparison to the  fuzzy ellipsoid}

\begin{figure}[tbph]
\center
\includegraphics*[width=150mm]{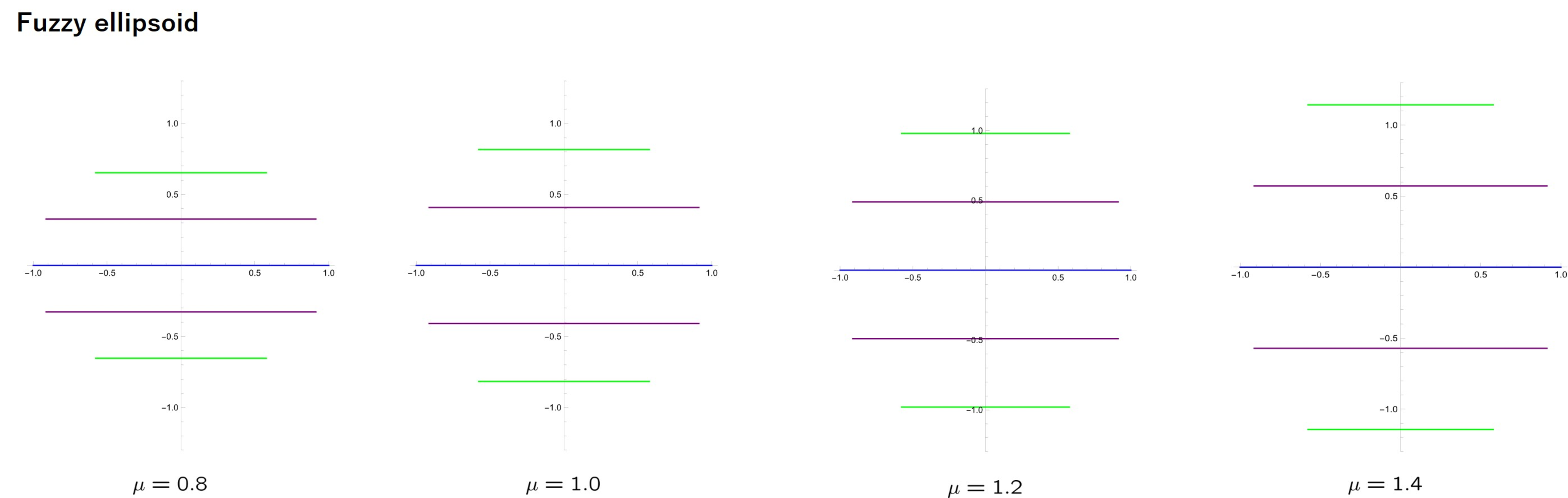}
\caption{Evolution of the fuzzy ellipsoid of $I=5$.    }
\label{ideal.fig}
\end{figure}

As a final subject, we investigate the difference between the fuzzy ellipsoid and the ellipsoidal matrix geometry.  
The fuzzy ellipsoid is defined as  
\be
{X'}^2+{Y'}^2+\frac{1}{\mu^2}{Z'}^2 =\frac{I-1}{I+1} \bs{1}_{I}. 
\label{fuzzydef}
\ee
The matrix coordinates of the fuzzy ellipsoid (\ref{fuzzydef}) are simply realized as  
\be
X'=\frac{2}{I+1} S_x^{(\frac{I}{2}-\frac{1}{2})}, ~~~Y'=\frac{2}{I+1} S_y^{(\frac{I}{2}-\frac{1}{2})}, ~~~Z'=\mu  \frac{2}{I+1} S_z^{(\frac{I}{2}-\frac{1}{2})}. 
\label{matrespco}
\ee
In Fig.\ref{ideal.fig}, we depict the evolution of  the normalized fuzzy ellipsoid coordinates. Comparing Fig.\ref{ideal.fig} with Fig.\ref{relafuzz.fig}, we can see that the evolution of the fuzzy ellipsoid is faster than that of the ellipsoidal matrix geometry. 
The fuzzy coordinates (\ref{matrespco}) satisfy 
 the $su(2)$ algebra 
\be
\!\!\!\![{X'}, {Y'}] =i\frac{2}{I+1}\frac{Z'}{ \mu}, ~~~~[{Y'}, \frac{Z'}{ \mu}] =i \frac{2}{I+1}{X'}, ~~~~[\frac{Z'}{\mu}, {X'}] =i\frac{2}{I+1} {Y'}. \label{fuzzyellipalga}
\ee
Equation (\ref{fuzzydef}) is invariant under the $SU(2)$ rotations generated by $X'$, $Y'$ and $\frac{1}{\mu}Z'$, which corresponds to the $SU(2)$ symmetry of the classical  ellipsoid (see Sec.\ref{subsec:classellip}). The fuzzy ellipsoid algebra (\ref{fuzzyellipalga}) is a linear algebra and  does not apparently correspond to a quantum version of the non-linear algebra of the classical  ellipsoid (\ref{poisellip}). Hence, the fuzzy ellipsoid  fails to realize the canonical  quantization of the ellipsoid. 

\begin{figure}[tbph]
\center
\includegraphics*[width=160mm]{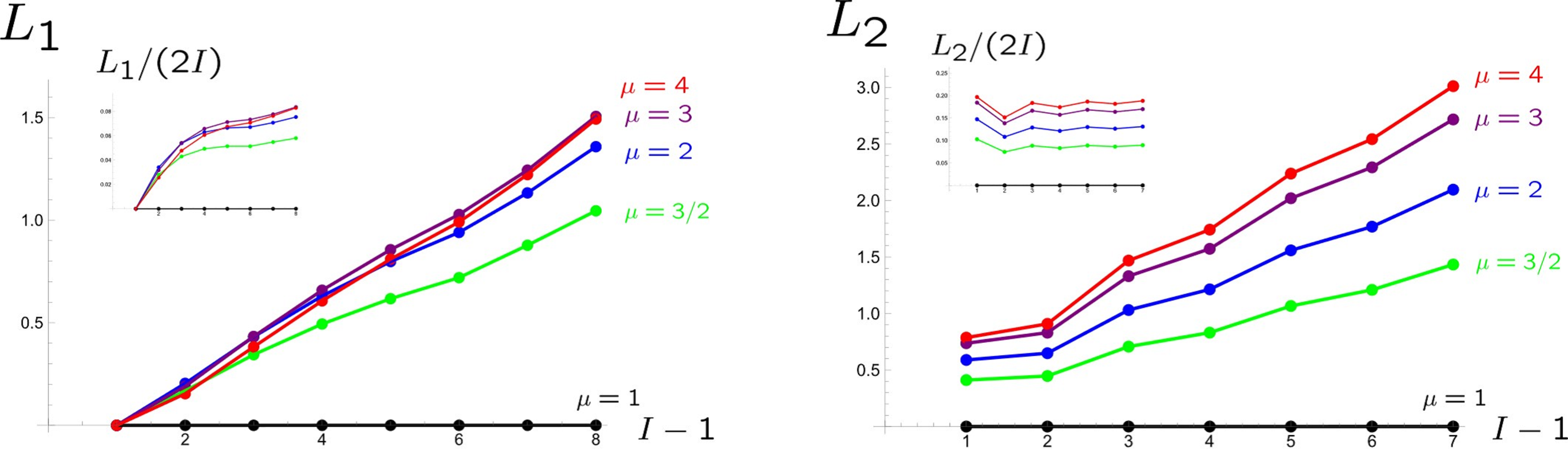}
\caption{Left: the behavior of $L_1$ (\ref{disunit}).    Right: the behavior of $L_2$ (\ref{distofelli}).  
At $\mu=1$, the ellipsoidal matrix geometry is reduced to the fuzzy sphere, so  $L_1$ and $L_2$ are identically zeros. For $\mu\neq 1$, $L_1$ and $L_2$ generally grow as $I$ and $\mu$ increase.  Insets:  the behaviors of the matrix distances normalized by $2I$. While the rates of increase   become milder, the data  imply that there still  exist non-vanishing finite distances between the ellipsoidal matrix geometries and the fuzzy ellipsoids even in the continuum limit $I\rightarrow \infty$. } 
\label{distance.fig} 
\end{figure}

While the  ellipsoidal matrix geometry only respects the $U(1)$ symmetry,  the fuzzy ellipsoid enjoys the $SU(2)$ symmetry.   
Since their symmetries are different,   the fuzzy ellipsoids and the ellipsoidal matrix geometries are   considered to describe different non-commutative geometries.  
We clarify quantitative differences between  the fuzzy ellipsoid and the ellipsoidal matrix geometry. Here, we consider the ellipsoidal matrix geometry of the  zeroth Landau level. 
We first evaluate the  Hilbert-Schmidt distance between ${X}^2+{Y}^2+\frac{1}{\mu^2}{Z}^2$ and $r(I-1)^2\bs{1}_I$ (\ref{rradform}) for the ellipsoidal matrix geometry: 
\be
L_1=\tr\biggl( \sqrt{\biggl({X}^2+{Y}^2+\frac{1}{\mu^2}{Z}^2-r(I-1)^2\bs{1}_I\biggr)^2  }~ \biggr). \label{disunit}
\ee
 If $X_i$ were the coordinates on the fuzzy ellipsoid $X'_i$, the quantity inside the trace  of (\ref{disunit}) would become zero (see Eq.(\ref{fuzzydef})). Then, the finite $L_1$ signifies  the difference  to the fuzzy ellipsoid. 
The behavior of $L_1$ is depicted in the left of Fig.\ref{distance.fig}, implying that the difference between the fuzzy ellipsoid and the ellipsoidal matrix geometry remains finite or  even  increases as the matrix size increases. 
As another  index for their difference, we  may adopt the following quantity:    
\be
L_2=\tr\biggl(\sqrt{(\hat{X}-\hat{X}')^2+(\hat{Y}-\hat{Y}')^2+\frac{1}{\mu^2}(\hat{Z}-\hat{Z}')^2}\biggr), \label{distofelli}
\ee
where  $\hat{X}_i$ and $\hat{X}_i'$ denote the normalized coordinates  of the ellipsoidal matrix geometry and those of the fuzzy ellipsoid,  respectively.\footnote{To reduce the overall scaling, we adopted  the normalized matrix coordinates (\ref{normamatco}).}  Obviously, $L_2$ increases as their  difference increases. 
The behavior of the $L_2$ is illustrated in the right of Fig.\ref{distance.fig},  which also implies  
  that the ellipsoidal matrix geometries do not coincide with the  fuzzy ellipsoids and  their difference even increases as  $\mu$ and $I$ increase.  

\section{Summary}\label{sec:summary}

 Exploring quantum matrix geometry of non-symmetric manifolds is  crucial for  further developments of  non-commutative geometry. In this paper, we proposed a new scheme for the construction  based on the  spectral flow, which naturally embodies the concept of the (time-)evolution of the matrix geometry.   
This scheme has the following properties: $(i)$  deriving matrix  coordinates is straightforward;   $(ii)$  non-perturbative deformation is available; $(iii)$  unitarity is preserved. Based on this scheme, we explicitly obtained the matrix geometries of the expanding sphere and ellipsoid  from the non-relativistic and relativistic Landau models.  
The magnetic field is chosen to be compatible with the  geometry of the base-manifold. 
  We also numerically confirmed the Atiyah-Singer index theorem  during the deformation from sphere to ellipsoid.  The ellipsoidal matrix geometries show qualitatively similar behaviors to the evolution  of the classical ellipsoid. The ellipsoidal matrix geometry uniquely evolves in   each of the  Landau levels.      
We confirmed that   the ellipsoidal matrix geometries are consistently reduced to the classical ellipsoids in the commutative limit.     We also illustrated  the differences between the ellipsoidal matrix geometry and the fuzzy ellipsoid.

Since this method is versatile, we can apply it to various manifolds including   higher dimensional manifolds and topologically non-trivial manifolds.   It may be  interesting to consider how the topological change of non-commutative space can be described in the present scheme.    
 According to the Atiyah-Singer index theorem, the topological number of the base-manifold is associated with the number of the zero-modes of the Dirac operator. Meanwhile, in the context of the matrix geometry, the number of the zero-modes is equal to the size of the matrix coordinates. Then, the topology change of the base-manifold necessarily varies the number of associated zero modes, leading to the change of the size of the non-commutative space. The topological change is thus intimately related to the violation of the unitarity of the non-commutative space.  
 While the present approach is advantageous for deriving an explicit  realization of the matrix coordinates, the underlying non-commutative algebra is difficult to specify. This issue should be  addressed in future research.

\section*{Acknowledgments}

The author would like to thank Goro Ishiki,  Naoki Sasakura, Kentaro Sato, and Asato Tsuchiya for fruitful discussions and Akifumi Sako for helpful email correspondence.  I am  also very grateful to Professor Shuang Zhang for his warm hospitality  at  the University of Hong Kong.  
 This work was supported by JSPS KAKENHI Grant No.~21K03542. 
 
\appendix

\section{Riemannian geometry of ellipsoid}\label{append:rieman}

We parametrize the ellipsoid 
\be
x^2+y^2+\biggl(\frac{z}{\mu}\biggr)^2=1, \label{ellipdefin}
\ee
as 
\be
x=\sin\theta\cos\phi, ~~~y=\sin\theta\sin\phi,~~z=\mu\cos\theta, 
\ee
where $\theta=[0, \pi]$ and $\phi=[0, 2\pi)$. 
From the formula of the  world line $ds^2=(\frac{\partial \bs{x}}{\partial\theta}d\theta +\frac{\partial \bs{x}}{\partial\phi}d\phi)^2$,  
the non-zero components of the metric are  derived as 
\be
g_{\theta\theta}=\rho^2, ~~~g_{\phi\phi}=\sin^2\theta,  \label{metr2com}
\ee
where 
\be
\rho =\sqrt{\cos^2\theta +\mu^2\sin^2\theta}. 
\ee
The zweibein is  derived as  
\be
e_1 =\rho~ d\theta, ~~~~~~e_2 =\sin\theta d\phi,  
\ee
and the spin-connection $\omega_{12}$ $(de_m+\sum_{n=1}^2\omega_{mn}e_n=0)$ is 
\be
\omega_{12} =-\omega_{21}=-\frac{1}{\rho} \cos\theta d\phi. \label{spinconel}
\ee
The corresponding curvature two-form is obtained as 
\be
R_{12}=d\omega_{12}  =\mu^2\frac{\sin\theta}{\rho^3}~d\theta\wedge d\phi, \label{2formel}
\ee
and its integral yields 
\be
\int R_{12} =4\pi. 
\ee
Using the metric (\ref{metr2com}),  the area element of the ellipsoid is derived as 
\be
d\Omega =  \sqrt{g}~d\theta  d\phi= \rho~\sin\theta d\theta  d\phi \label{areaele}
\ee
and the area of the ellipsoid $S=\int_{\text{ellipsoid}}d\Omega$ is calculated as 
\begin{subequations}
\begin{align}
&\mu>1 ~(\text{prolate spheroid})~~~:~S=2\pi \biggl(1+\frac{\mu^2}{\sqrt{\mu^2-1}}\text{Arcsec}(\mu)\biggr), \\
&\mu=1 ~(\text{sphere})~~~~~~~~~~~~~~~~:~S=4\pi , \\
&\mu<1 ~(\text{oblate spheroid})~~~~:~S=2\pi \biggl(1+\frac{\mu^2}{\sqrt{1-\mu^2}}\text{Arcsech}(\mu)\biggr). \label{areaexelliob}
\end{align}\label{areaexelli}
\end{subequations}
In the squashed-sphere-limit $\mu\rightarrow 0$, the oblate spheroid is completely flattened to be a double-sided unit disk with area $\pi$ for each. Equation (\ref{areaexelliob}) indeed yields    $S\overset{\mu \rightarrow 0}{\longrightarrow}  2\times \pi $. 
The Gauss curvature $K$  is derived as 
\be
K=\frac{1}{2}\mathcal{R}=\frac{\mu^2}{\rho^4},
\ee
which is related to the curvature two-form (\ref{2formel}) by  
\be
R_{12} =e_1 \wedge e_2 K. 
\ee
The Euler number is evaluated as 
\be
\chi =\frac{1}{2\pi}\int d\theta d\phi\sqrt{g}~ K =\frac{1}{2\pi}\int R_{12}=2. 
\ee

\section{Free particle on an ellipsoid}

For completeness, we present the spectral flow of the free particle quantum mechanics on an ellipsoid.

\subsection{Non-relativistic free particle on an ellipsoid}\label{sec:ellipsymm}

\begin{figure}[tbph]
\center
\includegraphics*[width=120mm]{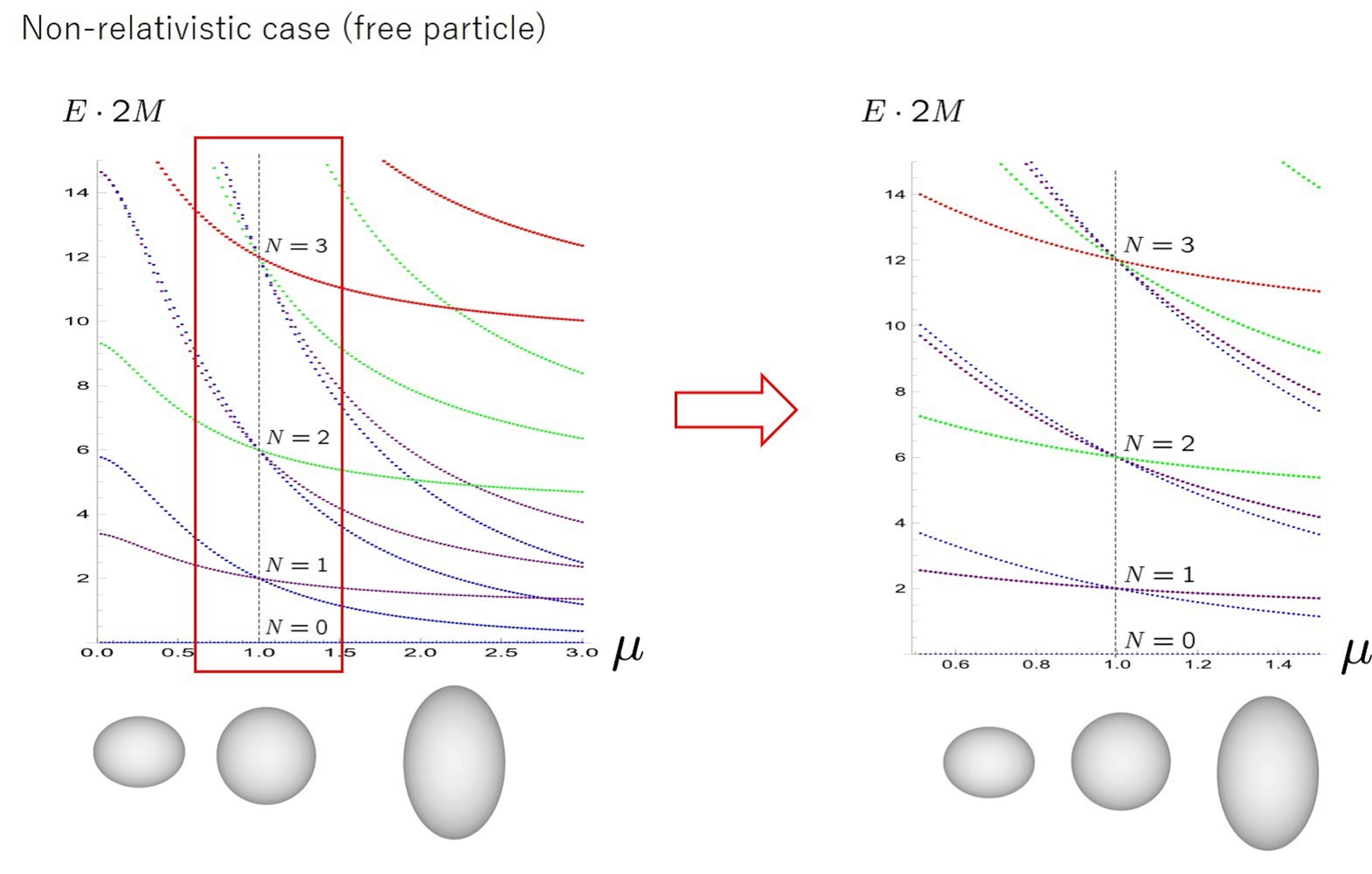}
\caption{The spectral flow of the eigenvalues of (\ref{freehamelli}).    At $\mu=1$ (sphere), the Hamiltonian has $SU(2)$ symmetry and its eigenvalues are given by $N(N+1)$ ($N=0,1,2,\cdots$) with degeneracy $2N+1$ and the corresponding eigenstates are the spherical harmonics.    The $m=0, \pm 1, \pm 2, \pm 3$  correspond to the blue, violet, green, and red curves, respectively.   The lowest energy remains at zero for arbitrary $\mu$.   }
\label{freeelli}
\end{figure}

The Schr\"odinger equation on an ellipsoid is given by\footnote{
At the disk limit $\mu~\rightarrow~0$,  (\ref{freehamelli}) is reduced to the polar coordinate Hamiltonian on a disk, 
\be
H_0 ~~\rightarrow~~-\frac{1}{2M}(\frac{\partial^2}{\partial \tilde{r}^2} +\frac{1}{\tilde{r}}\frac{\partial}{\partial \tilde{r}} +\frac{1}{\tilde{r}^2}\frac{\partial^2}{\partial \phi^2}), 
\ee
where $\tilde{r}:= \sin\theta$. 
}
\be
H_0=-\frac{1}{2M}\frac{1}{\sqrt{g}}\partial_{\mu}(\sqrt{g}g^{\mu\nu}\partial_{\nu})=-\frac{1}{2M} \biggl(\frac{1}{\rho^2}\partial_{\theta}^2 +\cot\theta \frac{1}{\rho^4}\partial_{\theta} +\frac{1}{ \sin^2\theta} {\partial_{\phi}}^2\biggr),
\label{freehamelli}
\ee
where $g_{\mu\nu}$ denote the metric on the ellipsoid (\ref{metr2com})  and 
\be
\rho := \sqrt{{\mu^2}(x^2+y^2) +\frac{1}{\mu^2} z^2} =\sqrt{\cos^2\theta +\mu^2 \sin^2\theta}. 
\ee
The quantum mechanical Hamiltonian (\ref{freehamelli}) does not  have the $SU(2)$ symmetry (except for $\mu=1$), but only has the $U(1)$ symmetry generated by $L_z$: 
\be
[H_0, L_z]=0. 
\ee
Obviously, the lowest energy of $H_0$ is zero for any value of $\mu$ and the lowest energy state is given by the constant wavefunction: 
\be
\varphi =\frac{1}{\sqrt{S}}, \label{constsdef}
\ee
where $S$ is the area of the ellipsoid (\ref{areaexelli}). The constant wavefunction (\ref{constsdef}) corresponds to the $s$-wave  spherical harmonics in the spherical system. The spectral flow of $H_0$ is depicted in Fig.\ref{freeelli}.

\subsection{Relativistic free particle on an ellipsoid}\label{sec:relellipsymm}

\begin{figure}[tbph]
\center
\includegraphics*[width=120mm]{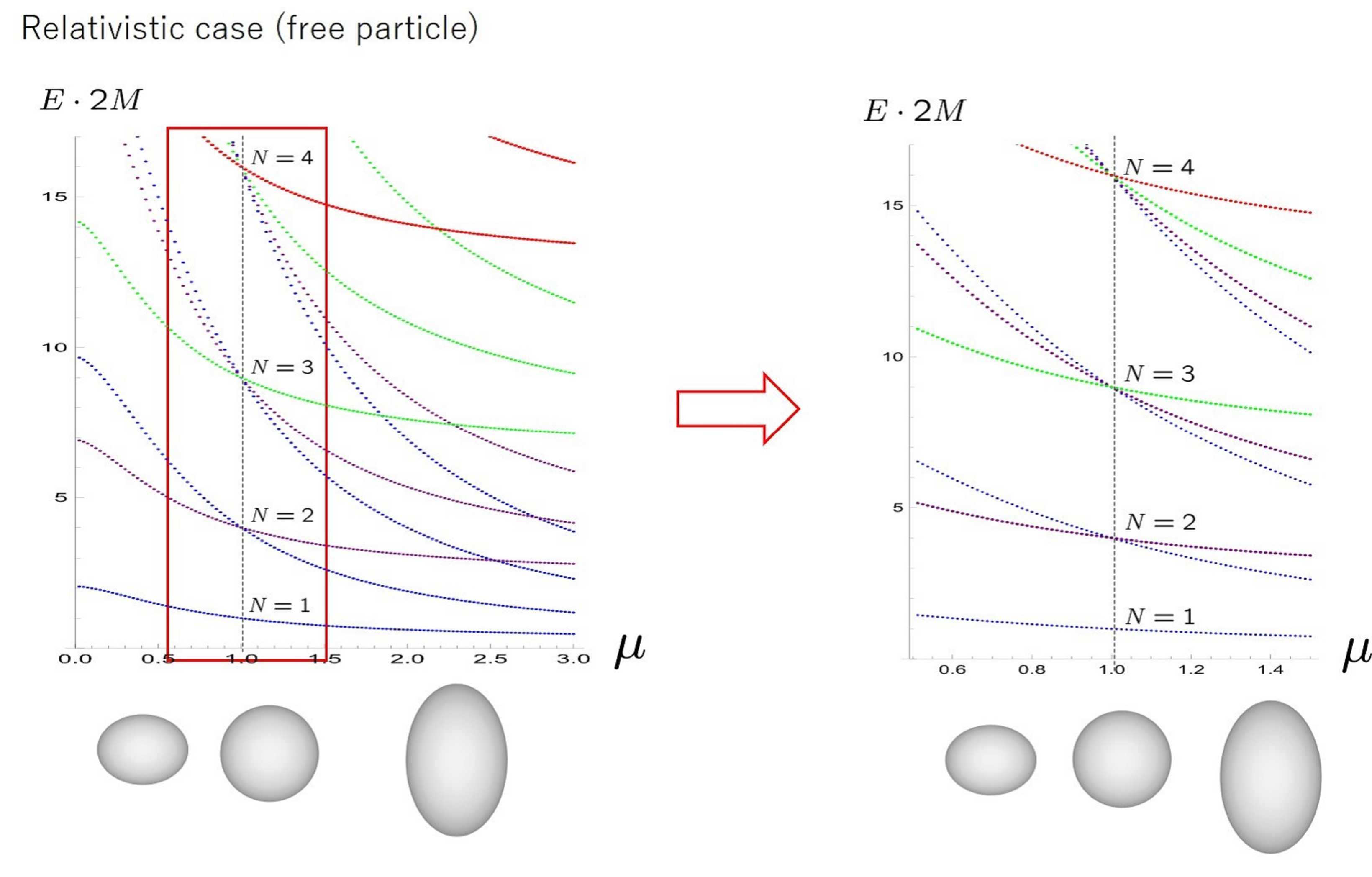}
\caption{The spectral flow of the eigenvalues of (\ref{freesqdex}).   At $\mu=1$ (sphere), the square of the Dirac operator has the eigenvalues $N^2$ ($N=1,2, 3, \cdots$) with  $4N$-fold degeneracy. The $m=\pm 1/2, \pm 3/2, \pm 5/2, \pm 7/2$  with spin degrees of freedom correspond to the blue, violet, green and red curves, respectively.   }
\label{relfreeelli}
\end{figure}

The square of the free Dirac operator is  represented by the following Lichnerowicz formula 
\be
(-i\fsl{{\nabla}})^2 =-\Delta +\frac{\mathcal{R}}{4}, \label{sqnabfree}
\ee
where $\mathcal{R}=2\frac{\mu^2}{\rho^4}$ denotes the scalar curvature and $\Delta$ signifies a Laplace-Beltrami operator,   
\be
\Delta := \frac{1}{\sqrt{g}} \mathcal{\nabla}_{\mu} (\sqrt{g}g^{\mu\nu}\mathcal{\nabla}_{\nu}).
\ee
With the spin connection (\ref{spinconel}), the covariant derives,  
$\mathcal{\nabla}_{\mu} := \partial_{\mu} +i\omega_{\mu}$,   
are given by 
\be
\mathcal{\nabla}_{\theta} =\partial_{\theta}, ~~~~~~\mathcal{\nabla}_{\phi} =\partial_{\phi} +i\omega_{\phi}=\partial_{\phi}-i\frac{1}{2\rho} \cos\theta \sigma_3,
\label{caldcov}
\ee
and  (\ref{sqnabfree}) is expressed as 
\be
(-i\fsl{\mathcal{\nabla}})^2 
=-\frac{1}{\rho^2 }\partial^2_{\theta}-\cot\theta \frac{1}{\rho^4}\partial_{\theta} -\frac{1}{\sin^2\theta}\mathcal{\nabla}_{\phi}^2 +\frac{1}{2}\frac{\mu^2}{\rho^4} . \label{freesqdex}
\ee
For $\mu=1$, (\ref{freesqdex}) is reduced to  
\be
(-i\fsl{\mathcal{\nabla}})^2|_{\mu=1} =-\partial^2_{\theta}-\cot\theta \partial_{\theta} -\frac{1}{\sin^2\theta}\biggl(\partial_{\phi}-i\frac{1}{2}\cos\theta \sigma_3\biggr)^2 +\frac{1}{2}. 
\ee
The spectral flow of (\ref{freesqdex}) is depicted in Fig.\ref{relfreeelli}. 

\end{document}